\documentstyle[12pt,epsfig]{article}

\begin{document}
\begin{center}

            \Large{\bf{I.JAVAKHISHVILI TBILISI STATE UNIVERSITY}}
\end{center}
\vspace{1.2cm}
\hspace{9.0cm}{\large\bf{\it{With the rights of the manuscript}}}
\vspace{1.2cm}
\begin{center}
\large{\bf{Tamar Djobava}}           
\end{center}
\vspace{1.cm}
\begin{center}

            \Large{\bf{Study of Multiparticle Production Processes
in Relativistic Nucleus-Nucleus Collisions by Using Streamer 
Spectrometer}}
\end{center}
\vspace{1.2cm}
\begin{center}

            {\bf{01.04.16- Nuclear and Elementary Particle Physics}}
\end{center}
\vspace{1.2cm}
\begin{center}

            \large{\bf{ A b s t r a c t $~~$ o f $~~$ D i s s e r t a t i o n}}
\end{center} 
\begin{center}

\large{\bf{Submitted in Fulfilment of Scientific Degree
of Doctor of Physical and Mathematical Sciencies}}
\end{center}
\vspace{3.5cm}
\begin{center}

            \Large{\bf{TBILISI, 2003}}
\end{center}
\newpage
{\underline{\bf{Subject of the Research Work.}}} 
Heavy ion reactions at high energies are able to create a strongly compressed and highly
excited nuclear matter in laboratory. At such extreme conditions a number of new
phenomena like phase transition of nuclear (hadronic) matter into the pion condensate,
 density isomers and a deconfined state	of nuclear matter -- the Quark-Gluon
Plasma (QGP) are predicted theoretically. In the QGP quarks and gluons are no longer
confined inside individual nucleons and mesons, but are free to wander over  distances
much longer than 1 fermi, the characteristic size of a hadron.
\par
All phenomena mentioned above deal with collective behaviour of the nuclear matter.
First the matter was observed to be flowing sidewards in the reaction plane due to the
high pressure developed at the impact. Later it was also seen that the matter is 
squeezed out of the hot zone between the two nuclei in the orthogonal direction to the
reaction plane. Collective flow effects in heavy ion collisions such as the
transverse (directed) and elliptic (squeeze-out) flows are expected to provide insight
into the properties of hot and dense nuclear matter and information about its Equation 
of State (EOS).
The transverse (directed) and elliptic flow effects have been already observed
for protons, light nuclei, pions, kaons, and $\Lambda$-hyperons in nucleus-nucleus collisions
at the energies of (0.1--1.8) GeV/nucleon at BEVALAC (Lawrence Berkeley Laboratory - USA),
GSI/SIS (Darmstadt, Germany), at 3.7 GeV/nucleon at Dubna (Joint Institute of Nuclear Research -
Russia), at 2--14 GeV/nucleon at AGS BNL (Brookhaven National Laboratory -- USA) and
at 60 and 200 GeV/nucleon at CERN SPS (European Organization for Nuclear Research -- Switzerland),
and more recently by the STAR and PHENIX collaborations at RHIC BNL at 
$\sqrt{s_{NN}}$=130 GeV/nucleon.
The elliptic flow of nucleons and pions was found to change its orientation from
out-of plane at 1 A$\cdot$GeV to in-plane at 11.6 A$\cdot$GeV. 
Many different methods were proposed for the study of  flow effects in relativistic 
nuclear collisions, of which most commonly used are the directed transverse momentum analysis 
technique developed by P.Danielewicz and G. Odyniec and the method of the Fourier expansion
of azimuthal distributions proposed by Demoulins et. al. and S. Voloshin and Y.Zhang.
\par
For an  interpration of the promising possible signatures of the quark-gluon plasma a better
understanding of the reaction dynamics is mandatory.
In order to investigate and understand the dynamics of relativistic nucleus-nucleus
collisions, it is important to have information on the rapidity and transverse momentum
distributions of nucleons participating in the interaction and of particles (mainly
pions) produced in the collisions, all as a function of the size of the system and
of the impact parameter. The transverse momentum and rapidity distributions provide 
information on the degree of stopping, thermalization, expansion and flow effects in
 the collisions. All of this information helps to obtain a consistent picture of
the entire reaction process.
The temperature and density of nuclear matter are among the main parameters of
the EOS determining the phase transition mechanism. The strongly interacting hadrons
are expected to decouple in the late stages of the collision. Their transverse
momentum spectra should therefore provide information about the conditions of the
system at freeze-out, in particular about  the temperature and collective velocity
of the system. To obtain the temperature of secondary hadrons in the experiment, one
usually estimates the inclusive spectrum slope.
\par
Pion production is the most important inelastic channel in relativistic heavy ion 
reactions. The yield of pions in the final state is connected, through
the population of intermediate resonances, to the temperature within the reaction zone.
Thus, although not directly related to the EOS, pions are an integral part of the effort
to determine  properties of nuclear matter. Because of their light masses and large
cross sections for interaction in nuclear matter, $\pi^{-}$ mesons are expected 
to thermalize easily. Furthermore, their spectra are not affected as those of heavier
particles by a given collective flow velocity. The multiplicity and transverse
momentum spectra of produced pions can provide information 
on properties of both the initial and
 final state properties of the hot hadronic matter. Thus, pions are 
a good probes for the study
of properties of nuclear matter in heavy ion collisions. Moreover, pions also influence
the production of the particles such as dileptons and high energy  photons which 
themselves have been used as probes of EOS.
\par
At high energies different dynamical mechanisms contribute to spectra of
secondaries. Among them "pionization" and fragmentation mechanisms are
widely discussed. "Pionization" means the existence of secondary pions with
relatively low momenta and flat (almost isotropic) angular distribution in
the centre of mass frame of colliding objects. The fragmentation
component has sharply anisotropic angular distribution in the centre of
mass frame. One of the principal problems in this direction is the
separation of these two components. Up to now there exists no unique way to
separate these mechanisms. Different authors propose different approaches and
non of them seems to be satisfactory.  The
presentation  of inclusive spectra in terms of light front variables
provides a possibility to separate these two components.\\
{\underline{Aims of the Thesis}:\\
$\bullet$
Study of properties of $\pi^{-}$ mesons produced in C-C, F-Mg and Mg-Mg
collisions at energy of 3.4 GeV/nucleon at a GIBS set-up of JINR/Dubna. Investigation
of multiplicity distributions of $\pi^{-}$ mesons, characteristics of these
distributions (the average multiplicity $<n_{-}>$, the dispersion of multiplicity
distribution $D_{n_{-}}$, etc),  average number of interating 
nucleons $<Q>$ and ratio $R_{-}= <n_{-}>/<Q>$ in central C-C, F-Mg and Mg-Mg 
interactions. Study of average kinematical characteristics of $\pi^{-}$ mesons   
such as momentum, transverse momentum, emission angle, rapidity and corresponding 
distributions in Mg-Mg collisions. Analysis of rapidity distributions
of pions in Mg-Mg ineratcions in various ranges of transverse momentum.\\
$\bullet$ 
Analysis of angular distributions of $\pi^{-}$ mesons in He-Li, He-C, C-Ne,
C-Cu, C-Pb, Mg-Mg and O-Pb collsions at energy of 3.7 GeV/nucleon obtained by use of 
SKM-200 and GIBS spectrometers. Extraction of the anisotropy coefficinet $a$.
Study of the dependence of the $a$ on mass numbers of the projectile ($A_{p}$)
 and target ($A_{T}$), on kinetic energy in c.m.s and multiplicity of pions $n_{-}$.\\
$\bullet$
Study of characteristics of protons in central He-Li, He-C, C-Ne, C-Cu and 
C-Pb collisions, namely the average values of momentum, transverse momentum,
 rapidity and emission angle.\\
$\bullet$
  Extraction  of temperatures of protons, $\pi^{-}$ and $\pi^{+}$ mesons 
from kinetic energy and transverse momentum ($P_{T}$) spectra in 
central He-Li, He-C, C-C, 
C-Ne, C-Cu , O-Pb and  Mg-Mg collisions. Study of the dependence of temperatures
of protons ($T_{p}$) and $\pi^{-}$ mesons ($T_{\pi^{-}}$) on the atomic numbers of 
projectiles and targets, rapidity and emission angle.
The investigation of a correlation of pion kinematical characteristics 
with $\Lambda$'s momentum
in the nucleon-nucleon  (N-N) c.m. system in
Mg-Mg collisions.\\
$\bullet$ 
Investigation of inclusive spectra of $\pi^{-}$ mesons in He(Li,C), C-Ne,
C-Cu, Mg-Mg and O-Pb collisions in terms of light front variables in order to
divide the phase space of secondary pions into two regions (parts)
with the aim to separate  "pionization" and fragmentation mechanisms. 
Study of 
characteristics of momentum, angular and $P^{2}_{T}$ distributions of pions in these
two regions with the aim to obtain information  whether the characteristics are 
similar or different in these two parts.\\
$\bullet$
  Investigation of the directed and elliptic  collective flow effects of protons and 
$\pi^{-}$ mesons in central C-Ne and C-Cu collisions at energy of 3.7 GeV/nucleon by
use of the transverse momentume technique  developed by P.Danielewicz and G.Odyniec. Study of 
dependence of flow parameter $F$ (the measure of the amount of collective transverse
momentum transfer in the raction plane) on the target mass number ($A_{T}$). Comparison
of $F$ with flow data for various projectile/target configurations at GSI-SIS, AGS and
SPS-CERN energies using the scaled flow $F_{s}=F/(A_{P}^{1/3}+A_{T}^{1/3})$. Extraction
of the parameter $a_{2}$ of the anisotropy term $a_{2}cos2\phi$ from  azimuthal 
distributions of protons and $\pi^{-}$ mesons with respect to the reaction plane at
mid-rapidity region in both C-Ne and C-Cu collisions. Study of the ratio $R$ of the
 number of particles emitted in the perpendicular direction to the number of particles
 emitted in the reaction plane, which represents the magnitude of the out-of-plane
emission signal. Study of dependence of $a_{2}$ and $R$ on target mass numbers, 
rapidity and transverse momentum.\\
$\bullet$ 
Measurement of the differential transverse flow of protons and $\pi^{-}$ mesons in
central C-Ne and C-Cu collisions. Investigation of the strength of
differential transverse flow of protons and
$\pi^{-}$ mesons .\\
$\bullet$
  Comparison of obtained experimental results with predictions of the Dubna
 Intranuclear Cascade Model (DICM), Quark Gluon String Model (QGSM) and Hagedorn
Thermodynamical Model.\\
{\underline{Scientific Novelty}}\\
$\bullet$
The experimental data has been obtained on SKM-200 and GIBS set-ups of JINR
by using the triggering system for selection of inelastic and central collisions,
which are characterized by a large  set of colliding pairs of nuclei
$A_{P}(A_{P}=^{4}He,~^{12}C,~^{16}O,~^{24}Mg) + 
A_{T}(A_{T}=^{6}Li,~^{12}C,~^{20}Ne,~^{24}Mg,~^{64}Cu,~^{207}Pb)$ at a momentum
of 4.3 and 4.5 GeV/c per nucleon, by the pure target-nuclei and also by different
impact parameters.\\
$\bullet$
 Properties of $\pi^{-}$ mesons produced in C-C, F-Mg and Mg-Mg
collisions at energy of 3.4 GeV/nucleon at a GIBS set-up of JINR/Dubna has been studied.
 Multiplicity distributions of $\pi^{-}$ mesons , characteristics of these
distributions (the average multiplicity $<n_{-}>$, the dispersion of multiplicity
distribution $D_{n_{-}}$, etc), 
the average number of interating 
nucleons $<Q>$ and ratio $R_{-}= <n_{-}>/<Q>$ 
interactions were investigated.
 Study of average kinematical characteristics of $\pi^{-}$ mesons   
such as momentum, transverse momentum, emission angle, rapidity and corresponding 
distributions in Mg-Mg collisions has been carried out.
 Analysis of rapidity distributions
of pions in Mg-Mg interactions in various ranges of transverse momentum was performed.\\
$\bullet$ 
 Angular distributions of $\pi^{-}$ mesons in He-Li, He-C, C-Ne,
C-Cu, C-Pb, Mg-Mg and O-Pb collsions at energy of 3.7 Gev/nucleon obtained by use of 
SKM-200 and GIBS spectrometers has been analysed.
The anisotropy coefficient $a$ have been extracted. 
Study of the dependence of the $a$ on mass numbers of the projectile ($A_{p}$)
 and target ($A_{T}$), on kinetic energy in c.m.s and multiplicity of pions $n_{-}$
was performed.\\
$\bullet$
  Characteristics of protons  was studied in central He-Li, He-C, C-Ne, C-Cu and 
C-Pb collisions, namely the average values of momentum, transverse momentum,
 rapidity and emission angle.\\
$\bullet$
 Temperatures of $\pi^{-}$ and $\pi^{+}$ mesons 
 has been estimated
from kinetic energy and transverse momentum ($P_{T}$) spectra in 
central He-Li, He-C, C-C, 
C-Ne, C-Cu , O-Pb and  Mg-Mg collisions for the first time.
 The dependence of temperatures
of protons ($T_{p}$) and $\pi^{-}$ mesons ($T_{\pi^{-}}$) on the atomic numbers of 
projectiles and targets, rapidity and emission angle has been investigated.\\
$\bullet$ 
Inclusive spectra of $\pi^{-}$ mesons in He(Li,C), C-Ne,
C-Cu, Mg-Mg and O-Pb collisions in terms of light front variables
 $\xi^{\pm}$ and $\zeta^{\pm}$ have been investigated for the first time.
In  $\zeta^{\pm}$ ($\xi^{\pm}$) distributions  the points
$\tilde{\zeta^{\pm}}$ ($\tilde{\xi^{\pm}}$) have been singled out, which
divide the phase space of secondary
$\pi^{-}$ mesons into two regions
with significantly different characteristics,
 in one of which the thermal
equilibrium seems to be reached.
 Characteristics of $\pi^{-}$ mesons (the momentum, angular,
$p_{T}^{2}$ -- distributions) in these two regions differ significantly.\\
$\bullet$
  The directed and elliptic  collective flow effects of protons and 
$\pi^{-}$ mesons in central C-Ne and C-Cu collisions at energy of 3.7 GeV/nucleon 
has been obtained  for the first time by
use of transverse momentume technique developed by  P.Danielewicz and G.Odyniec.  
Dependence of flow parameter $F$ (the measure of the amount of collective transverse
momentum transfer in the raction plane) on the target mass numbers ($A_{T}$)
was studied. Comparison
of $F$ with flow data for various projectile/target configurations at GSI-SIS, AGS and
SPS-CERN energies using the scaled flow $F_{s}=F/(A_{P}^{1/3}+A_{T}^{1/3})$
has been carried out.
 The parameter $a$ of the anisotropy term $a_{2}cos2\phi$ from  azimuthal 
distributions of protons and $\pi^{-}$ mesons with respect to the reaction plane at
mid-rapidity region in both C-Ne and C-Cu collisions has been extracted
for the first time.
 The ratio $R$ of the
 number of particles emitted in the perpendicular direction to the number of particles
 emitted in the reaction plane, which represents the magnitude of the out-of-plane
emission signal was studied. The dependence of $a_{2}$ and $R$ on target mass numbers, 
rapidity and transverse momentum has been investigated.
The second Fourier coefficient $v_{2}=<cos2\phi>$, which is related to
$a_{2}$ via $v_{2}=a_{2}/2$ and measures the elliptic flow, have been estimated
both for C-Ne and C-Cu. The obtained results of the  elliptic flow
excitation function for protons in C-Ne and C-Cu collisions 
has been compared with the data
 in the available energy region of 0.2 $\div$
200.0 GeV/nucleon. 
The excitation function $v_{2}$  clearly
shows an evolution from negative ($v_{2} < 0$) to positive ($v_{2} > 0$)
elliptic flow within the region of $2.0 \leq E_{beam} \leq 8.0$ GeV/nucleon
and point to an apparent transition energy $E_{tr} \sim 4$ GeV/nucleon.
The obtained collective flow effects show the persistence of these
collective phenomena in the range of available energy, i.e. from the
Bevalac and GSI/SIS up to Dubna, AGS, RHIC and SPS energies.\\
$\bullet$ 
 The differential transverse flow of protons and $\pi^{-}$ mesons in
central C-Ne and C-Cu collisions have been measured for the first time. 
 The strength of protons and
$\pi^{-}$ mesons differential transverse flow has been investigated.\\
$\bullet$
 Obtained experimental results have been compared
with predictions of Dubna
 Intranuclear Cascade Model (DICM), the Quark Gluon String Model (QGSM)
and Hagedorn Thermodynamical Model.
The models satisfactorily describe different experimental results.\\
{\underline{Practical Value}}\\
$\bullet$
 Statistical programs of experimental data processing and analysis have been
elaborated in High Energy Physics Institute of
Tbilisi State University, which can be used for the processing and analysis of
the data of the other experiments.\\
$\bullet$
 Obtained experimental results can be used for planning and performing of
new experimental investigations of nucleus-nucleus interactions at high energies and
are important  for checking-up (verifing) of the theoretical models, which
give predictions about nucleus-nucleus collisions in detail.\\ 
{\underline{Structure and Basic Contents of the Thesis.}
The thesis is based on  results obtained on SKM-200 and GIBS set-ups.
 Results of investigation enriched the  assumptions about the dynamics of 
nucleus-nucleus collisions.  
\par
The dissertation contains an introduction, 6 chapters, conclusions and
 referred bibliography. It consists of 251 pages, accounts 88 figures and
18 tables. The references include 212 items.
\par
The experimental data obtained at the Laboratory of High Energies (LHE)
 of JINR/Dubna by the SKM-200-GIBS International Collaboration (Alma-Ata - Moscow - Dubna
 - Warshawa - Tbilisi) have been processed and analysed at LHE of JINR and
at the High Energy Physics Institute of Tbilisi State University.
\par
{\underline{Introduction}} presents the subjects of the thesis, its aims and
obtained results. Structure of the thesis and its short overview are given.
\par
{\underline{Chapter 1}} is devoted to description of the spectrometers SKM-200 and
 GIBS which have been constructed at the Laboratory of High Energies at JINR, Dubna.
 Spectrometers consist of a 2m streamer chamber placed in a magnetic field of 
$\sim$ 0.8 T ($\sim$ 0.9 T for GIBS) and
 a triggering system.  Main elements  of both spectrometers and the logic of
 triggering system are described and the physical parameters of spectrometers
are compared. The streamer chamber was exposed to beams of $^{4}He$, $^{12}C$,
$^{16}O$, $^{20}Ne$ (SKM-200) and $^{12}C$, $^{19}F$, $^{24}Mg$ (GIBS) nuclei
accelerated in the synchrophasotron up to the energy of 3.7 GeV per incident nucleon.
\par
Solid targets Li, C, Cu and Pb in the form of thin discs with thickness of
(0.2 $\div$ 0.5) g/cm$^{2}$ (the thickness of Li was 1.5  g/cm$^{2}$) were mounted
within the fiducial volume of the SKM-200 chamber. Neon-gass filling of the chamber,
also served as a nuclear target. C, F, and Mg solid targets with thickness of
0.99 g/cm$^{2}$, 1.34 g/cm$^{2}$ and 1.56 g/cm$^{2}$, respectively, were
used for GIBS set-up.  
 The "inelastic" trigger, consisting of two sets of scintillation
counters mounted upstream and downstream
from the chamber, was selecting all inelastic interactions of incident nuclei with a target.
The "central" triggering system was selecting events defined as those without charged
projectile spectator fragments and spectator neutrons ($P/Z >$ 3 GeV/c), 
emitted within a cone of half angle $\Theta_{ch}$=$2.4^{0}$ or $2.9^{0}$
and  $\Theta_{n}$=$1.8^{0}$ or $2.9^{0}$ for SKM-200, and 
$\Theta_{ch}$= $\Theta_{n}$=$2.4^{0}$ for GIBS set-up.
\par   
The streamer chamber pictures were scanned twice at HEPI TSU and JINR/Dubna.  
Final results of scanning and measurement of events have been recorded on 
the Data Summary Tapes (DST) and then have been analysed by using the standard 
statistical programs HBOOK, PAW, etc.
The data processing and analysis procedures are 
described in detail in the Chapter 1.
\par
In {\underline{Chapter 2}} three models, namely
the Dubna version of Intranuclear Cascade Model (DICM), the Quark Gluon String 
Model (QGSM) and Hagedorn Thermodynamic model, are described which
have been used for
comparison with the experimental results presented in the dissertation.
In the Intranuclear Cascade model each of colliding nuclei are represented as the Fermi
gas of nucleons concluded in the potential well and particles are produced
 in the independent hadron-hadron collisions. 
 Collision dynamic is traced in time by use of the Monte-Carlo technique. The
 model takes into account the simultaneous development of intranuclear cascades
in both interacting nuclei. Mathematically, it is described by the relativistic
Boltzmann equation for the one-particle distribution function  of a mixture of gases.
In the DICM the effect of Coulomb forces
acting between the projectile and target nuclei is taken into account approximately.
\par
The Quark-Gluon String Model (QGSM) allows to investigate hadron-nucleon, hadron-nucleus
and nucleus-nucleus collisions at wide energy interval and especially in the range of
 intermediate energy ($\sqrt{s} \leq 4$ GeV).
The Quark-Gluon String Models are the next generation of cascade models and are based on 
 the formation of coloured field tubes or quark-gluon
strings ( the excited objects, consisted of quarks connected via gluon string) 
during the collision process, which decay into the observable hadrons.
The QGSM is based on
the Regge and string phenomenology of particle production in inelastic
binary hadron collisions. To describe the evolution of the hadron and
quark-gluon phases, the model uses a coupled system of Boltzmann-like
kinetic equations. The Monte-Carlo solution of these system of equations corresponds
to the direct modelling  procedure of collisions.
The procedure of event generation consists of 3 steps: 1. definition of
configuration of colliding nucleons, 2. production of quark-gluon strings
and 3. fragmentation (breakup) of strings  into observed hadrons.
The coordinates of nucleons are generated according to a realistic
nuclear density.
The sphere-nuclei are filled with the nucleons at a
condition that the distance between them is greater than 0.8 fm. Nucleon
momenta are distributed in the range of 0 $\leq$ P $\leq$ $P_{F}$.
The maximum nucleon Fermi momentum is the following:
\begin{eqnarray}
 P_{F}=(3\pi^{2})^{1/3}h \rho(r)^{1/3} 
\end{eqnarray}
 where h=0.197 fm$\cdot$GeV/c. For main NN and $\pi$N
interactions the following topological quark
diagrams were used: binary, 'undeveloped' cylindrical, diffractive and 
planar. Binary process makes a main contribution.
It corresponds to quark rearrangement without direct particle emission in
the string decay. This reaction predominantly results in the production of
resonances (for instance, $ N + N$$\rightarrow$$ N + $$\Delta^{++}$), 
which are the main source of pions. 
The QGSM simplifies the nuclear effects. In particular, coupling of nucleons
inside the nucleus is neglected, and the decay of excited recoil nuclear fragments
and coalescence of nucleons are not included.
\par
The statistical bootstrap  model (SBM) of Hagedorn  represents 
the generalization of the thermodynamic two-fireball model. In the Hagedorn model
instead of two fireballs,  the infinite number of fireballs is considered, which
are moving toward the c.m.system with different Lorentz factors. In the model 
two simple mechanisms have been used, the thermodynamics and relativistic kinematics
and with the help of these mechanisms a large  number of experimental
data have been described.
In the thermodynamic approach the production of particles is assumed to be  due to 
geometry and available volume in the space phase. 
\par
{\underline{Chpater 3}} is devoted to a detailed study of pion production in central nucleus-
nucleus collisions at energy of 3.4 $\div$ 3.7 GeV/nucleon. The advantage
of $\pi^{-}$ mesons choice with respect to other particles in order
to establish characteristics of collision dynamics, is shown. 
 Besides, production of pions is the
predominant production process at Dubna energies.
\par
 Discussion of possible sources
of experimental biases and appropriate correction procedures is given.
  Sources of systematic errors are: admixture of nuclei with $Z < Z_{proj}$,
trigger bias, collisions with gaseous $^{14}N$, depletion of narrow events,
secondary interactions within solid targets, low momentum pions absorption, admixture
of $e^{-}$,  $K^{-}$ and $\Sigma^{-}$ and scanning losses.
Inaccuracy of corrections was calculated, and systematic
uncertainties of these corrections were found to be less than statistical
ones. The total
uncertainty of corrections depending on the trigger,
material and  the thickness of target has been estimated and was about
 $(1.5 \div 4.5) \%$.
\par
Results of study of 
 multiplicity distributions of $\pi^{-}$ mesons, characteristics
of these distributions (average multiplicity $<n_{-}>$, dispersion  $D_{n}$,
 etc), the average number of interacting nucleons
$<Q>$, the ratio $R_{-}=<n_{-}>/<Q>$, the dependence of $<Q>$ and $R$ on  masses of
colliding nuclei in central C-C, F-Mg and Mg-Mg interactions at energy of
 3.4 GeV/nucleon are presented in Chapter 3.
 Results were obtained by scanning of streamer
chamber pictures. Pictures were scanned
twice, and a subsequent third scan was used to resolve ambiguities or discrepancies
between the first two scans. Efficiency of scanning, defined as the ratio
of the average number of particles obtained after the single scan to the
average number of paricles after the third scanning, turned out to be equal
to 0.987 $\pm$ 0.013. $\pi^{-}$ mesons with $P > 50$ MeV/c have been registered.
 In order
to reduce the systematic errors to the minimal
values, caused both by secondary interactions
within the solid target and by screening of the  part
 of the event by the passing beam tracks, and to obtain the final result,
from initial samples of events ( 1583 C-C, 1557 F-Mg and
6239  central Mg-Mg collisions) 
 have been chosen the events with the
following conditions:
a) $N_{q} \leq Z_{p}+Z_{t}$, where $N_{q}=M-2n_{-}$, $M$ - is the total number
of observed charged particles and $n_{-}$ - is the number of negative pions;
b) only one beam track enters the chamber.
 Finally obtained subsamples, called further the "refined"
("cleaned") events, formed about $70 \%$ from the initial ones.
 Besides  average multiplicities and dispersions of multiplicity
distributions of pions in C-C, F-Mg and Mg-Mg collisions 
the average numbers of interacting nucleons (protons) $<Q>$ and the ratios
$R=<n_{-}>/<Q>$ have been obtained also. The value of $<Q>$  is defined by the formula:
\begin{eqnarray}
<Q> =<M>- 2<n_{-}> - <n_{S}> - <n_{b}>
\end{eqnarray}
where $<M>$ is the average number of all observed charged particles (excluding
the identified $e^{+}$ and $e^{-}$), $<n_{-}>$ - the average number of negative pions,
$<n_{S}>$ and $<n_{b}>$ --  average numbers of charged spectator fragments of
projectile and target, respectively. $<M>$ and $<n_{-}>$ have been defined by
the scanning results. 
 The term $<n_{b}>$ in (2) cuts off the evaporated part of the spectra of
charged fragments, emitted from the target. Due to the convention
of the definition of evaporated particle and in order to check the stability of
results,
for all targets two cut
off levels have been determined : $P/Z$ = 240 and 310 MeV/c, ( which for
protons correspond to $E_{kin}$=30 and 50 MeV). As the difference
in results obtained at two above mentioned limits does not exceed
the statistical error, further have been presented the averaged values, which
correspond to the level of cut off $P/Z \simeq $ 280 MeV/c ( for protons
$E_{kin} \simeq $ 40 MeV).
Limits, which define the stripping particles (spectator fragments of the projectile),
 namely: the emission angle $\Theta < 4 ^{0}$ and
$P/Z >$ 3 GeV/c in the rest frame of the projectile -  approximately correspond
to the upper energy limit ($E_{kin}$=50 MeV) for $b$ particles, emitted from
the target. The contamination of evaporated (or stripping) particles with
$Z > $ 1 does not affect $<Q>$, as in (2) they enter with equal weight
in $<M>$ and $<n_{b}>$ (or $<n_{S}>$).
\par
Values of $<Q>$ obatined in  this way from (2) correspond to the number
of interacting protons of projectile and target
with some admixture of heavier partciles.  Systematic
error of $<Q>$ due to the admixture of particles with $Z$=2 is
estimeated to be $\leq 1 \%$.
 The dependence of $<Q>$ on the target mass number $A_{T}$ have been studied. 
For this purpose our result for C-C has been compared to 
the data of C-Ne, C-Cu, C-Zr and C-Pb collisions for the trigger
T(2,0) obtained in our experiment at SKM-200 set-up earlier.
The $<Q>$ increases with $A_{T}$.  The data have been approximated
with the empirical dependence $C \cdot A_{T}^{\alpha}$, where the following
values for the parameters $C$ and $\alpha$ have been obtained :
$C=2.65 \pm 0.15$, $\alpha=0.48 \pm 0.02$. 
 The Dubna version of Cascade Model (DICM) satisfactorily 
describes the data.
\par
The ratio $R_{-}=<n_{-}>/<Q>$ have been studied intensively at Berkeley
experiments and in our experiment earlier. This ratio is
interesting as in the number of models the pion multiplicity normalized
on the number of interacting nucleons is calculated more simply then the total
multiplicity. The calculation is especially simplified at $A_{P}=A_{T}$, since
in this case the center mass system   
does not depend on the impact parameter $b$ in the collective
models due to the symmetry of the colliding parts of the nuclei. 
Therefore, calculations for central collisions usually are
carried out at $b$=0, avoiding the intergration over $b$.
\par
The ratio $x=A_{P}/A_{T}$ have been used for the quantitative analysis
of the $R_{-}$ dependence on  the atomic numbers of colliding nuclei.
 Fig. 1 presents the dependence of $R_{-}$ on $A_{P}/A_{T}$.
  In Fig. 1 along with our result for C-C, F-Mg and Mg-Mg collisions 
the data of C-$A_{T}$, O-$A_{T}$  and Ne-$A_{T}$
 collisions for the trigger
T(2,0) obtained in our experiment at SKM-200 set-up earlier, are presented as well.
One can see, that $R_{-}$ depends on $x=A_{P}/A_{T}$, increasing with
$A_{P}/A_{T}$. The data have been approximated (the solid line)
with empirical dependence $C \cdot (A_{P}/A_{T})^{\alpha}$, where the following
values for parameters $C$ and $\alpha$ have been obtained :
$C=0.49 \pm 0.02$, $\alpha=0.22 \pm 0.02$. Crosses in Fig. 1 are the result
of the calculation by the Dubna version of Intranuclear Cascade Model (DICM). One can see,
that the model satisfactorily describes the data: the discrepancy with
the model does not exceed $20 \%$.
\par
Values of $R_{-}$ obtained for C-C, F-Mg and Mg-Mg
interactions ($R_{-}$= 0.47$\pm$0.04, $R_{-}$= 0.45$\pm$0.03 
and $R_{-}$= 0.46$\pm$0.03,
respectively) coincide with results obtained 
earlier at SKM-200 set up
for symmetric and approximately symmetric systems of C-Ne, O-Ne and Ne-Ne
($R_{-}$= 0.44$\pm$0.03, $R_{-}$= 0.50$\pm$0.04 and $R_{-}$= 0.41$\pm$0.05,
respectively).
The Propane Chamber collaboration at Dubna obtained
the following values of $<Q>$ and $R_{-}$ in  central C-C collisions
at energy of 3.4 GeV/nucleon:
$<Q>=8.92 \pm 0.05$, $R_{-}=0.35 \pm 0.01$.  The value
of $<Q>$ coincides with our result $<Q>=8.80 \pm 0.50$
 and the disagreement in $R_{-}$
may be the consequence of systematic error in
$<n_{-}>$ values of Propane collaboration for C-C collisions.
 In the energy interval of E=0.4 $\div$ 1.8 GeV/nucleon for
central Ar-KCl collisions it have been obtained that
 the $R_{-}$ increases linearly with
$E_{cm}$:
\begin{eqnarray}
R_{-}(E_{cm}) =\frac{<n_{-}(E_{cm})>}{<Q(E=1.8)>}=0.74 \cdot (E_{cm}-0.1)
\end{eqnarray}
Extrapolating (3) to our energy ($E_{cm}$=0.673 GeV/nucleon), one can
obtain $R_{-}$=0.42, which coincides with our result and extends the
linear dependence of $R_{-}$ for symmetric pairs of nuclei, observed in Bevalac
experiments, up to our energy of 3.6 GeV/nucleon and
 allows one to extend the region of comparison with the predictions
of theory beyond the limits of Bevalac  (LBL) energies.
Results of $R_{-}$ for symmetric pairs of nuclei C-C, Mg-Mg
are compared with the calculations of simple
thermodynamic model, which takes into account the pionic degrees of freedom.
 The  obtained values of $R_{-}$ coincide with the values calculated by using
the thermodynamical model at $\rho / \rho_{0} \sim 2$, where
 $\rho$  
is the nuclear density at freeze-out temperature $T$ and $\rho_{0}$ is the
normal (ground state) nuclear density.
\par
 A study of pion production in central Mg-Mg collisions is described in Chapter 3.
 The average kinematical characteristics of $\pi^{-}$ mesons
 such as multiplicity, momentum,
transverse momentum, emission angle, rapidity,
and   corresponding distributions have been obtained. Experimental results
 have been compared with predictions of Quark-Gluon String Model.
 Mg-Mg  interactions  have been generated  using Monte-Carlo generator
COLLI, which is based on the QGSM.
 Events had been traced through the detector and trigger filter.
Events had been generated for not fixed impact parameter $\tilde{b}$.
  From the impact parameter distribution
we obtained the mean value of
$<b>$=1.34 fm.
For the obtained value of $<b>$, we have generated
a total sample of 6200 events. 
The results obtained by the model in two regimes (for $\tilde{b}$ and 
for b=1.34 fm)
are consistent and it seems, that in our experiment the value of
b=1.34 fm for Mg-Mg is most probable.
 From the analysis of generated events
the pions with deep angles greater than 60$^{0}$ had been excluded, because
efficiency of registration of such vertical tracks  in the experiment 
is low. Comparison of $n_{-}$, $P$, $\Theta$ , $P_{T}$ and
$Y$ distributions allows
to conclude, that the QGSM satisfactorily describes the spectra and
experimental and model values of average kinematical characteristics
coincide within the errors.
\par
Two Lorenz invariant variables have been used to describe the main features of the
$\pi^{-}$ mesons, produced in nucleus-nucleus collisions: the rapidity Y and
the transverse momentum P$_{T}$.
For this purpose we investigated rapidity
distributions in various regions of P$_{T}$: P$_{T}$ $\leq$ 0.2,
0.2 $\leq$ P$_{T}$$<$ 0.3,
0.3 $\leq$ P$_{T}$$<$ 0.5, P$_{T}$ $\geq$ 0.5. These distributions
have the characteristic Gaussian form.
 Shape of $Y$
distributions changes with increase of the transverse momentum
of $\pi^{-}$ mesons: the fraction of pions increases in the central
region and decreases in fragmentational regions of colliding nuclei.
 Analysis of $Y$ distributions showes, that the central regions
of these distributions are enriched with  pions of large transverse
momentum  (as compared to fragmentational regions of colliding nuclei):
$\langle$$P_{T}$$\rangle$ =$0.184\pm0.004$ GeV/c  for  Y$<$ 0.2,
$\langle$$P_{T}$$\rangle$ =$0.247\pm0.002$ GeV/c  for
0.7$\leq$Y$\leq$1.6,
$\langle$$P_{T}$$\rangle$ =$0.189\pm0.003$ GeV/c  for  Y$>$ 2.
The QGSM reproduces the $Y$ distributions in the various regions
of  P$_{T}$ quite well. 
Practically all theoretical models are based on the dependence of average
kinematical characteristics on thicknesses of colliding nuclear layers.
For the fixed $A_{T}$ the thickness of nuclear matter is connected with
the impact parameter {\it{b}}. It is, therefore, desirable to study the dependence
of  kinematical characteristics of pions on the impact parameter and on mass
numbers of colliding nuclei ($A_{P}$, $A_{T}$). As {\it{b}} is
experimentally unmeasurable, its estimation can be  obtained
on the basis of the number
of nucleons of the projectile participating in  the interaction
$\nu_{p}$, which in turn is
correlated with the multiplicity of observed $\pi^{-}$ mesons n$_{-}$.
Therefore, the study of dependence of a given variable (P$_{T}$, $Y$, etc)
on {\it{b}} can be qualitatively replaced by the study of the dependence
of the same variable on $n_{-}/A_{P}$. The variable $n_{-}/A_{P}$ 
 serves as the measure of the impact parameter at fixed $A_{T}$.
The Quark Gluon String model  allows us to obtain the distribution of the
impact parameter {\it{b}} for any given trigger mode and any $n_{-}/A_{P}$
values. From the QGSM data of Mg-Mg collisions, it has been obtained, 
 that the average impact parameter is strongly
correlated with $n_{-}/A_{P}$. Thus any observed dependence on the trigger
mode or $n_{-}/A_{P}$ can be translated into a dependence on the $<\it{b}>$
value.
Therefore, in a qualitative analysis one can investigate the dependence of
kinematical characteristics on $n_{-}/A_{P}$.
A quantitative analysis of the shape of $P_{T}$ and $Y$ distributions
have been carried out based on a study of the statistical moments
of the distributions, namely the average values ($<P_{T}>$ and $<Y>$ )
and the dispersions ($D_{P_{T}}$, $D_{Y}$).
 It has been shown, that  in Mg-Mg collisions $<Y>$ does not depend
on the multiplicity of pions, meanwhile
$<P_{T}>$ slightly
decreases with multiplicity. These average kinematical characteristics
are similar to the characteristics of N-N collisions at the same energy.
 Obtained results for Mg-Mg interactions
agree  with our previous results for symmetric and approximately
 symmetric pairs of nuclei C-C, C-Ne, O-Ne, Ne-Ne.
 The QGSM reproduces the dependence of
$<Y>$ and $<P_{T}>$ on $n_{-}$ for Mg-Mg. 
\par
 Analysis of angular distributions of $\pi^{-}$ mesons  in   He-Li,
He-C, C-Ne, C-Cu,  C-Pb,  O-Pb  and  Mg-Mg  collisions is presented in Chapter 3.
 The anisotropy coefficient {\it {a}}  has been   obtained from the
cos$\Theta$$^{*}$  distributions  of $\pi^{-}$ mesons in c.m.s, approximating them by
 the ansatz:
\begin{eqnarray}
     dN/dcos\Theta^{*} =const (1 + {\it{a}} cos^{2}\Theta^{*})
\end{eqnarray}
 It  has
been shown, that the parameter {\it{a}}  is similar for the symmetric  (or
approximately symmetric, $A_{P}$$\sim$$A_{T}$ ) system of nuclei ( He-Li
and  Mg- Mg) and increases slowly with $A_{P}$  and $A_{T}$  for other pairs of
nuclei.
 The dependence of {\it{a}} on kinetic energy in c.m.s $E^{*}_{kin}$
 and multiplicity of pions  n$_{-}$ has been studied.
It has been shown, that the anisotropy
coefficient increases linearly with  E$^{*}$$ /$E$^{*}_{max}$ for all pairs of
nuclei. In  the range of 100 MeV  pions are emitted isotropically.
 Dependence of the
dispersion D$_{cos\Theta^{*}}$ on E$^{*}$$ /$E$^{*}_{max}$ is similar to the
dependence of the coefficient {\it{a}} on E$^{*}$$/ $E$^{*}_{max}$ .
Increase  of  the dispersion  D$_{cos\Theta^{*}}$ with E$^{*}$$
/$E$^{*}_{max}$   is well reproduced by the DICM.
At  small multiplicities of pions ( $n_{-} \leq~ <n_{-}>$ ) , the degree of
anisotropy is larger, than at high multiplicities  ( $n_{-} >~ <n_{-}>$ ) .
 Decrease of the parameter {\it{a}}  for more central events  ( $n_{-} >
~<n_{-}>$ ) indicates, that the angular distributions of pions become
more  isotropic for more central collisions (small impact parameters).
 The QGSM reproduces the angular distributions of pions well enough and
the values of the  anisotropy coefficient
{\it {a}}, extracted from the spectra generated by QGSM,  agree
with the exprimental ones within the errors.
\par
{\underline{Chapter 4}} 
is devoted to the analysis of the proton and $\pi^{-}$ meson
inclusive kinetic energy $E_{K}$ and transverse momentum $P_{T}$
spectra in central interactions of He, C and Mg nuclei with Li, C, Ne, Mg,
Cu and Pb nuclei with the aim to determine their temperatures.
One of the main topics of the current relativistic heavy ion experiments
is the determination of the properties of nuclear matter at high densities
and temperatures.
Such canonical parameters of the equation of state of nuclear matter (EOS),
as temperature, density, pressure may be experimentally deduced from the
system decay products alone. To obtain the temperature of secondary hadrons
in the experiment one usually estimates the value of the inclusive spectrum
slope.
\par
 Characteristics of the protons in central He-Li, He-C, C-C, C-Ne,
C-Cu and C-Pb collisions have been studied.
  For this purpose, the identification of protons have been carried out
in central C-Ne and C-Cu collisions.
The statistical method of identification utilized in the neural nets
method and based on the similarity of $\pi^{-}$ and $\pi^{+}$ mesons spectra
have been used in order to separate $\pi^{+}$ mesons  from protons.
The admixture of
$\pi^{+}$ mesons amongst the positive charged  particles is about
(25-27) $\%$. After identification the admixture of $\pi^{+}$ mesons have been reduced to
(5-7)$\%$.
After identification of $\pi^{+}$ mesons in C-Ne and C-Cu collisions,
proton distributions have been obtained. 
 For inelastic collisions of He,
all charged secondaries were measured and central subsamples T(2,0) were
selected. In He-Li, He-C, C-C and C-Pb collisions proton 
distributions have been obtained
by subtraction of inclusive $\pi^{-}$ meson  distributions from 
inclusive positive charged particle distributions, which
contain protons (with admixture of deuterons $d$ and tritons $t$) and
$\pi^{+}$ mesons, (the proton
mass was assigned to all charged particles). 
 The fact that in isospin symmetric
 nucleus-nucleus collisions the $\pi^{+}$ and $\pi^{-}$ distributions
are similar has been used.
 The average values of
momentum, transverse momentum, rapidity and emission
angle have been obtained. Experimental results have been compared
with the predictions of the Quark-Gluon String Model. 
The model
satisfactorily describes the main characteristics and spectra of the
protons, though  slightly underestimates $<P>$
and overestimates $<\Theta>$.
\par 
 Temperatures of protons in He-Li, He-C, C-C, C-Ne, C-Cu and C-Pb
central collisions have been determined.
The proton and pion temperatures has been estimated by means of :
1) inclusive kinetic energy $E_{K}$ spectra and
2) transverse momentum $P_{T}$ spectra.
\par
Non-invariant inclusive spectra $d^{3}\sigma/dp^{3}=(E^{*}P^{*})^{-1}
dN/dE_{K}$ ($P^{*}$ is the momentunm, $E^{*}$ the total energy and
$E_{K}$ the kinetic energy of the particle in the c.m.s.) have been
analysed. A simple exponential law was used to fit each experimental
spectrum:
\begin{eqnarray}
F(E_{K})=(E^{*}P^{*})^{-1}dN/dE_{K}=A \cdot exp(-E_{K}/T)
\end{eqnarray}
$T$ is related to average kinetic energy of a given type of particles
and, thus, characterizes the nuclear matter temperature at that expansion
stage, when such particles are emitted. Therefore, the $T$ parameter is
usually called an average or inclusive temperature.
It has been state, that transverse momentum distributions are preferable
because of their Lorenz invariance during longitudinal boosts. Transverse
momentum distributions were described by the following form:
\begin{eqnarray}
dN/dP_{T}=const \cdot P_{T} \cdot E_{T} \cdot K_{1}(E_{T}/T)\simeq const \cdot
 P_{T} \cdot (TE_{T})^{1/2} \cdot exp(E_{T}/T)
\end{eqnarray}
\begin{eqnarray}
E_{T}=(P^{2}_{T}+m^{2})^{1/2}
\end{eqnarray}
$K_{1}(X)$ is Mac-Donald's function.\\
Temperatures  has been determined from the noninvariant kinetic energy ($E_{K}$) and
transverse momentum distributions of protons in the rapidity region of
$0.4 \leq Y \leq 1.6$ for He-Li, He-C, C-C and C-Ne collisions and
$0.3 \leq Y \leq 1.7$ for C-Cu and C-Pb.
Proton temperatures
$T_{p}$ increase with atomic numbers of projectile ($A_{P}$) and target
($A_{T}$) from $T_{p}=(118\pm2)$ MeV for He-Li to
$T_{p}=(141\pm2)$ MeV for C-Pb. 
 Results obtained from the experimental data have been compared with the
QGSM predictions. 
 The QGSM slightly overestimates proton
temperature  for He-Li and for other pairs of nuclei gives values consistent
with the experiment.
The QGSM data show  also the
similar  tendency for $T_{p}$ to increase with
the mass numbers of projectile ($A_{P}$) and target ($A_{T}$)
as experimental ones.
Hagedorn thermodynamical model at our energy predicts proton
temperatures $T_{p}=(135 \div 138)$ MeV. From our results only proton
temperatures in C-Ne (131$\pm$1 MeV), C-Cu (135$\pm$2 MeV) and C-Pb (141$\pm$4 MeV) 
 collisions agree with the model predictions.
The dependence of $T_{p}$ on the bombarding energy
$E_{lab}$ for different pairs of nuclei from a series of experiments of Bevalac (LBL),
GSI/SIS, Dubna (including our results), AGS and CERN/SPS has been studied.
\par
Temperatures of $\pi^{-}$  mesons  has been determined from the noninvariant 
kinetic energy ($E_{K}$) and
transverse momentum distributions in the
rapidity region of
$0.5 \leq Y \leq 2.1$ for light nuclei pairs of He-Li, He-C and C-Ne. 
 A single exponential fit  describes the pion spectra and $T_{\pi^{-}}$ does
 not depend on $A_{P}$, $A_{T}$.
The $T_{\pi^{-}}$ deduced from energy spectra is equal to
$T_{\pi^{-}}\simeq$ 87 MeV and from transverse momentum spectra to
$T_{\pi^{-}}\simeq$ 95 MeV.
 Results obtained from the experimental data have been compared with the
QGSM data. 
 Distributions of QGSM data satisfactorily describe the experimental
spectra and the QGSM gives consistent
with experiment values of pion temperatures for He-Li, He-C and C-Ne.
 The pion experimental spectra
in medium and heavy nuclei Mg-Mg, C-Cu, C-Pb (Fig. 2) and O-Pb  show a concave
shape  which can not be described by single exponential
law and a sum of two exponentials should be used to reproduce the data (with two
temperatures $T_{1}$ and $T_{2}$).
We have estimated the fraction $R_{2}$ of the pion yield that falls in the second
exponent, determined by $T_{2}$. Values of $R_{2}$ are equal to: $R_{2}=0.12\pm0.03$
for O-Pb , $R_{2}=0.22\pm0.02$ for Mg-Mg, $R_{2}=0.24\pm0.02$ for C-Pb and
$R_{2}=0.25\pm0.02$ for C-Cu.
The QGSM reproduces the spectra of C-Cu, C-Pb and Mg-Mg interactions and
the values of temperatures coinside with the experimental ones, only
for Mg-Mg the value of $T_{2}$ is slightly underestimated.
At our energy, the Hagedorn model predicts only one temperature
$T_{\pi^{-}}=(115 \div 120)$ MeV, which agrees within the errors
with our values of $T_{2}$.
 For C-Ne and C-Cu interactions the temperatures of identified
$\pi^{+}$ mesons  have been determined. One temperature have been observed
in C-Ne and two temperatures in C-Cu collisions, similarly as for the
$\pi^{-}$ mesons. 
\par
The concave shape of pion energy and transverse momentum
spectra in central heavy ion collisions
has been observed in many different experiments starting 
from Berkley and GSI energies up to AGS and CERN/SPS energies.
 This phenomena has been observed:
in Ar-KCl collisions at a beam energy of 1.8 GeV/nucleon,
La-La at E=1.35 GeV/nucleon and Au-Au at E=1.15 GeV/nucleon at
Berkeley, in Au-Au and Ni-Ni collisions at E=1$\div$2 GeV/nucleon at GSI
by FOPI collaboration at midrapidity
 ( -0.1$<y^{0}<$0, where y$^{0}$=y/y$_{cm}$-1),
in C-C and C-Ta collisions at E=3.4 GeV/nucleon
at Dubna by Propane Bubble Chamber collaboration, in Au-Au collisions
at E=10$\div$14 GeV/nucleon at AGS by E877, E866 collaborations. In
ultrarelativistic heavy ion collisions the  pion transverse momentum spectra
also show a concave shape.
Several hypothesis have been proposed in order to explain the concave shape of the
spectra. These include the superposition of thermal pions and pions
from the final-state $\Delta$ decays, higher resonances and the effect
of baryons flow on the pions. Based on an equilibrium model calculations,
it was also supposed, that the concave shape of the pion spectra may come
from an isotropic hydrodynamical expansion of the hot compressed nuclear matter.
The experimentally observed concave shape of the pion spectra is well reproduced
within the framework of a hadronic transport model of B.A Li and W. Bauer.
This model is based on solutions of a set of coupled transport equations for the
phase-space distribution functions of nucleons, baryon resonances ($\Delta$,
$N^{*}$), and pions. 
\par
Dependence of the $T_{\pi^{-}}$ on emission angle in c.m. system
$\Theta_{cm}$ have been studied in He-Li, He-C and C-Ne collisions.
The $T_{\pi^{-}}$ falls off from
$T_{\pi^{-}} \simeq 110$ MeV at $\Theta_{cm}=30^{0}$ to
$T_{\pi^{-}} \simeq 90$ MeV at $\Theta_{cm}=90^{0}$  and then increases.
 The QGSM reproduces this dependence. Similar dependence have been obtained
for Ar-KCl collisions at energy of E=1.8 GeV/nucleon. Such behaviour have been
interpreted as the effect of direct "corona" production of $\Delta$'s where
the $\Delta$ escapes without further rescattering and subsequently decays into
$N\pi$. This mechanism enhances the forward/backward yield of pions.
The rapidity dependence of the 
$T_{\pi^{-}}$ shows a bell-shaped behaviour with the maximum at Y=Y$_{beam}$/2.
 The dependence on the rapidity is more
evident for $T_{2}$ than for $T_{1}$ in Mg-Mg and C-Cu collisions . 
The QGSM reproduces the experimental results.
 Similar result have been observed
 for d-Ta, He-Ta, C-Ta inelastic
collisions at energy of 3.4 GeV/nucleon by  the Propane Bubble Chamber
collaboration of JINR and 
by FOPI collaboration at GSI in Ni-Ni collisions at E=1$\div$2 GeV/nucleon.
\par
The pion energy
spectra are quite different from the corresponding proton spectra.
The pion spectra exhibit the concave shape, whereas the proton spectra
are convex shaped. This difference was first interpreted as due to the
different source sizes at freeze-out time, later  it was realized
 that pions mainly originate from the decay of the $\Delta$ resonance,
and that their spectral shape is strongly influenced by the decay kinematics. 
\par
 A correlation of pion kinematical characteristics with $\Lambda$'s momentum
in the nucleon-nucleon  (N-N) c.m. system was investigated in
Mg-Mg collisions. Events with a $\Lambda$ produced within ("$\Lambda^{in}$" events)
and beyond ("$\Lambda^{out}$" events) the N-N c.m. kinematical limits for
nucleon-nucleon collisions were considered.
 Kinematical characteristics and temperatures of pions from
$\Lambda^{in}$  and $\Lambda^{out}$  events do not reveal any significant
difference when compared between and with corresponding characteristics and temperatures
of pions produced in ordinary central Mg-Mg collisions.
The QGSM reproduces
the experimental results and reveal similar  to experimental data trend.
\par
In {\underline{Chapter 5}} the light front variables $\zeta^{\pm}$ and $\xi^{\pm}$, which define 
the so called
horospherical coordinate system in the Lobachevsky space, have been used
to study inclusive spectra of $\pi^{-}$ mesons in He(Li,C), C-Ne,
C-Cu, Mg-Mg and O-Pb collisions.
An important role in establishing of many properties of multiple particle
 production plays  the choice of kinematic variables in terms of which observable
quantities are presented. The
variables which are commonly used are the following:
 the Feynman
$x^{}_F=2p^{}_z/\sqrt{s}$,
 rapidity $y={1\over{2}}{\rm ln}
[(E+p^{}_z)/(E-p^{}_z)]$,
transverse scaling variable $x^{}_T=2p^{}_T/\sqrt{s}$, etc.
Unified scale invariant variables $\xi^{\pm}$  for the presentation of single
particle inclusive distributions have been proposed,
which are defined  
in the centre of mass frame as follows:
\begin{eqnarray}
\xi^{\pm}&=&\pm {E\pm p_z\over{\sqrt{s}}}=\pm {E+|p_z|\over{
\sqrt{s}}}
\end{eqnarray}
where $s$ is the usual Mandelstam variable,
$E=\sqrt{p^{2}_z+p^{2}_T+m^{2}}$ and $p_{z}$ are
the energy and the $z$ - component of the momentum of produced particle.
The upper sign in Eq. (8) is used for the right hand side hemisphere and
the lower sign for the left hand side hemisphere.
In order to enlarge the scale in the region of small $\xi^{\pm}$,
 it is convenient also to
introduce the variables
\begin{eqnarray}
\zeta^{\pm}=\mp{\rm ln}|\xi^{\pm}|
\end{eqnarray}
In the limits of high $p_{z}$ ($|p_z|\gg p_T$)
and high $p_{T}$  ($p_T\gg |p_z|$) the $\xi^{\pm}$
variables go over to the well known variables $x_{F}$ and $x_{T}$. 
The principal differences of $\xi^{\pm}$ distributions as compared to the
corresponding $x_{F}$ -- distributions are the
following:
(1) existence of some forbidden region around the point $\xi^{\pm}=0$;
(2) existence of maxima at some $\tilde{\xi^{\pm}}$ in the region of
 relatively small $|\xi^{\pm}|$.
 3) existence of the limits for
$\vert\xi^{\pm}\vert \leq m/\sqrt{s}$.
The
maxima at $\tilde{\zeta}^{\pm}$ are also observed in the invariant
distributions $(1/\pi) \cdot dN/d\zeta^{\pm}$ (Fig. 3).
 However, the region
$|\xi^{\pm}|>|\tilde{\xi}^{\pm}|$ goes over to the region
$|\zeta^{\pm}|<|\tilde{\zeta}^{\pm}|$ and vice
versa (see Eqs. (8) and (9)).
The  value of maxima are observed
 at $\tilde{\zeta^{\pm}}=2.0\pm0.1$ for all pairs of nuclei.
The $\tilde{\zeta^{\pm}}$ is the function of the energy (see Eqs. (8), (9)) and
does not depend on the mass numbers of the projectile ($A_{P}$) and
target ($A_{T}$).
\par
In order to study the nature of these maxima we have divided the
phase space into two regions
$|\zeta^{\pm}|>|\tilde{\zeta}^{\pm}|$ 
and
$|\zeta^{\pm}|<|\tilde{\zeta}^{\pm}|$
and studied the
$p_{T}^{2}$ and the angular distributions of $\pi^{-}$ mesons in these regions
separately.
The numbers of pions in these two regions are approximately
equal. For example in C-Cu interactions in the region
$|\zeta^{\pm}|>|\tilde{\zeta}^{\pm}|$ the number of pions is equal to
--1987 and in
$|\zeta^{\pm}|<|\tilde{\zeta}^{\pm}|$ --- 2212.
The
$p_{T}^{2}$ and the angular distributions of $\pi^{-}$ mesons differ significantly in
$\zeta^{+} > \tilde{\zeta^{+}}$  and $\zeta^{+} < \tilde{\zeta^{+}}$
regions. The angular distribution of pions in the  region $\zeta^{+} <
\tilde{\zeta^{+}}$  (Fig. 4)  is  sharply  anisotropic in 
contrast to the almost flat distribution
in the region $\zeta^{+} > \tilde{\zeta^{+}}$ (Fig. 4).
The flat behaviour of the angular distribution allows one to think that
one observes a partial thermal equilibrium in the region
 $|\zeta^{\pm}| > |\tilde{\zeta^{\pm}}|$ ($|\xi^{\pm}| < |\tilde{\xi^{\pm}}|$)
of phase space.
The slopes of
$p_{T}^{2}$ -- distributions differ greatly in different regions of
$\zeta^{\pm}$. For example  in Mg-Mg interactions:
$<p_{T}^{2}>=(0.027\pm0.002$) (GeV/c)$^2$ in the region  $\zeta^{+} > \tilde{\zeta^{+}}$;
$<p_{T}^{2}>=(0.103\pm0.009$)  (GeV/c)$^2$ in the region
      $\zeta^{+} < \tilde{\zeta^{+}}$. Thus the values of $\tilde{\zeta^{\pm}}$
are the boundaries of the two regions with significantly
different characteristics of $\pi^{-}$ mesons. 
To describe the spectra in the region $\zeta^{+} > \tilde{\zeta^{+}}$ the
Boltzmann
\begin{eqnarray*}
  f(E)\sim e^{-E/T}
\end{eqnarray*}
distribution has been used.
The distributions
$1/\pi \cdot  dN/d\zeta^{+}$, $dN/dp_{T}^{2}$, $dN/dcos\Theta$
have been fitted by Boltzmann distribution in the region $\zeta^{+} >\tilde{\zeta^{+}}$
and the parameter $T$ is obtained by fitting the data. 
In order to determine how the characteristics vary the analysis
has been carried out also for $\tilde{\zeta^{+}}$=1.9 and 2.1.
The results are similar, but the joint fit of the distributions is
better for $\tilde{\zeta^{+}}$=2.0 (presented on Fig. 4).
\par
 The spectra of $\pi^{-}$ mesons in the region $\zeta^{+} >
\tilde{\zeta^{+}}$ are
satisfactorily described by the formulae which follow from
the thermal equilibration. The same formulae when extrapolated to the region
$\zeta^{+} < \tilde{\zeta^{+}}$
deviate significantly from the data. Therefore
in the region $\zeta^{+} < \tilde{\zeta^{+}}$ the $p_{T}^{2}$ -- distributions
has been fitted by the formula
\begin{eqnarray}
\frac {dN}{dp_{T}^{2}} \sim \alpha \cdot e^{-\beta_{1}P_{T}^{2}} +
(1-\alpha) \cdot e^{-\beta_{2}p_{T}^{2}}
\end{eqnarray}
and the $\zeta^{+}$ -- distributions by the formula
\begin{eqnarray}
\frac{1}{\pi}\cdot\frac {dN}{d\zeta^{+}} \sim (1 - \xi^{+})^{n}=
(1 - e^{-\vert \zeta^{+}\vert})^{n}
\end{eqnarray}
which is an analogue of the $(1-x_{F})^{n}$ dependence -- the result of
the well-known quark-parton model consideration, which for
$\pi^{-}$ mesons gives the value n=3.  The dependence
$(1 - e^{-\vert \zeta^{+}\vert})^{n}$ is in good agreement with experiment
in the region $\zeta^{+} < \tilde{\zeta^{+}}$ and deviates from it in the
region $\zeta^{+} > \tilde{\zeta^{+}}$.
Thus in the $\zeta^{\pm}$ ($\xi^{\pm}$) distributions we have singled out
points $\tilde{\zeta^{\pm}}$ ($\tilde{\xi^{\pm}}$) which separate in the phase
space two groups of particles with significantly different characteristics.
There are no such points in the $x_{F}$ and $y$ -distributions. 
 The dependence of $T$ on $(A_{P}*A_{T})^{1/2}$  has been studied. It have been 
obtained, that temperature decreases with the increasing of $(A_{P}*A_{T})^{1/2}$
i.e. with increasing number of participating nucleons.
 The experimental results have been compared with the predictions of QGSM.
The QGSM satisfactorily reproduces the experimental data for light and
intermediate-mass nuclei.
\par
The similar analysis of $\pi^{-}$ meson spectra produced in p-C, He-C, C-C
and C-Ta interactions at a momentum of 4.2 GeV/c/nucleon in the 2-metre
 Propane Bubble Chamber of JINR (Dubna)
has been carried out in light-front variables. 
The results obtained by Propane collaboration coincide with ours.
\par
{\underline{Chapter 6}} is devoted to the study of collective flow effects of protons and
$\pi^{-}$ mesons in central C-Ne and C-Cu interations
at a momentum of 4.5 GeV/c/nucleon (E=3.7 GeV/nucleon).
Two different signatures of collective flow have been predicted:
a) the bounce off of compressed matter in the reaction plane ( a sidewards
deflection of the spectator fragments - "bounce off", as well as the directed flow
of nucleons from the overlap region between the colliding nuclei (participants)
in the reaction plane - "side splash"), called the sideward
(also often termed directed) flow.
b) the squeeze-out of the participant matter out of the reaction plane
--  the elliptic flow.
Collective effects lead to characteristic, azimuthally asymmetric sidewards
emission of the reaction products. While the transverse flow in the reaction
plane is influenced by the cold matter deflected by the overlap region of
the colliding nuclei, the squeeze out is caused by the hot and compressed
matter from the interaction region which preferentially escapes in the
direction perpendicular to the reaction plane unhindered by the presence of
the projectile and target spectators.
A strong dependence of these collective effects on the nuclear equation
of state (EOS) was predicted. Due to its direct dependence on the
EOS, $P(\rho, S)$, flow excitation functions can provide unique information
about phase transitions: the formation of abnormal nuclear matter, e.g.
yields a reduction of the collective flow. A directed flow excitation
function as signature of the phase transition into Quark Gluon Plasma
(QGP)
has been proposed by several authors.
The knowledge of EOS is of fundamental interest and is also essential for
understanding of astrophysical phenomena such as the supernova explosions,
the properties of the core of compact stars (neutron stars), the evolution
of the early Universe and the formation of elements in stellar nucleosynthesis.
\par
Many different methods were proposed for the study of flow effects in
relativistic nuclear collisions, of which the most commonly used are 
 the transverse momentum analysis technique developed by
P.Danielewicz and G.Odyniec and Fourier analysis of the azimuthal distributions
on the event-by-event basis in relatively narrow rapidity windows  proposed  
by M.Demoulins, S.Voloshin and Y.Zhang. We have used the technique of P. Danielewicz
and G.Odyniec in our analysis. The method involves two basic
ideas:
1) to select the rapidity range and rapidity dependent waiting factors in
the center of mass system, which provide the reaction plane closest to
the real reaction plane, and
2) to remove trivial and spurious self-correlations from the projections.
The reaction plane is defined by the transverse vector
$\overrightarrow{Q}$
\begin{eqnarray}
\overrightarrow{Q_{j}}=\sum\limits_{i\not=j}\limits^{n}\omega_{i}
\overrightarrow{P_{{\perp}i}}
\end{eqnarray}
where $i$ is a particle index and $\omega_{i}$ is a weight. Pions are
not included. The reaction plane is the plane containing $\overrightarrow{Q}$
and the beam axis. The weight factor $\omega_{i}$, depends on the
rapidity of the emitted particle $i$, so that the central
rapidity region, where the particle emission is azimuthally symmetric,
is omitted, and the forward and backward rapidity regions get weight with
opposite signs: $\omega_{i}$ is taken as 1 for
 y$_{i}$$>$ y$_{c}$ , -1 for y$_{i}$$<$ y$_{c}$ and $\omega_{i}$=0
for -$y_{c}<y_{i}< y_{c}$,
where the cutoff rapidity y$_{c}$ is usually chosen to be
y$_{c} \approx 0.3$y$_{c}^{beam}$;
  y$_{i}$ is the rapidity of
particle $i$. This choice leads to the result that the forward and backward
moving particles, which are azimuthally anticorrelated if there is a
collective transverse flow, will contribute equally to
$\overrightarrow{Q}$. Autocorrelations are removed by calculating
$\overrightarrow{Q}$ individually for each particle without including that
particle into sum.
\par
The transverse momentum of each particle in the estimated reaction plane is 
calculated as:
\begin{eqnarray}
P_{xj}\hspace{0.01cm}^{\prime} = \frac{
\sum\limits_{i\not=j} \omega_{i} \cdot (
\overrightarrow{P_{{\perp}j}} \cdot \overrightarrow{P_{{\perp}i}})}
{\vert\overrightarrow{Q_{j}}\vert}
\end{eqnarray}
The reaction plane 
have been defined
for the participant protons i.e. protons which are not fragments
of the projectile ($P/Z >3$ GeV/c , $\Theta < 4^{0}$) and target
($P/Z <0.2$ GeV/c). They represent the protons participating in the collision.
As we study an asymmetric pair of nuclei, we chose to bypass the
difficulties associated with the center-of-mass determination and carried
out the analysis in the laboratory frame. The
original weight $\omega_{i}$ have been replaced by the continuous function
$\omega_{i}$= $y_{i}$ - $<y>$ as in papers of Bevalac streamer chamber
collaboration,
 where  $<y>$ is the
average rapidity, calculated for each event over all the participant protons.
The value of the weight $\omega_{i}$ should be chosen to minimize the
fluctuations of $\overrightarrow{Q_{j}}$ from the true reaction plane.
\par
 It is known, that the estimated reaction plane
differs from the true  one, in particular,
due to the finite multiplicity.
The component $ P_{x}$ in the true reaction plane is systematically larger
then the component $P_{x}\hspace{0.01cm}^{\prime}$
in the estimated plane. The correction factor $k$ is equal to
$k_{corr}$=1 $/$ $< cos\Delta\phi >$, where $\phi$ is the angle
between the estimated and true planes.
$k$ is subject to a large uncertainty,
especially for low multiplicity.
According to the method of Danielewicz, for the definition 
of $< cos\Delta\phi >$ we divided
randomly each event into two equal sub-events,
constructed vectors $\overrightarrow{Q_{1}}$ and
$\overrightarrow{Q_{2}}$ and estimated the azimuth angle difference
 $\phi_{12}$
between these two vectors. $< cos\Delta\phi > $ = $< cos\phi_{12} >$.
The data did not allow to perform the analysis for
different multiplicity
intervals, therefore we defined the correction factors $k_{corr}$, averaged
over all the multiplicities. The values of $k_{corr}$ are: $k_{corr}$=1.27$\pm$0.08
for C-Ne and $k_{corr}$=1.31$\pm$0.04 for C-Cu.
For the estimation of $< cos\Delta\phi > $
 the alternative method have been applied also, which does
not require the division of each event into two sub-classes.
\begin{eqnarray}
< cos\Delta\phi >  \approx <\omega P_{x}\hspace{0.01cm}^{\prime} >[<W^{2}-W>/
<Q^{2}-\sum (\omega_{i} P_{{\perp}i})^{2}>]^{1/2}
\end{eqnarray}
where $W$=$\sum \vert\omega_{i} \vert$. These two methods yield consistent
results within the errors.
Fig. 5  shows the dependence of the estimated $<P_{x}\hspace{0.01cm}^{\prime}(Y)>$
on $Y$ for protons
in C-Ne collisions. The average transverse momentum
$<P_{x}\hspace{0.01cm}^{\prime}(Y)>$ is obtained by averaging over
all events in the corresponding intervals of rapidity.
The data exhibit the typical $S$-shape behaviour
which demonstrates the collective transverse momentum transfer between the
forward and backward hemispheres. The directed flow is an odd function
of the rapidity. It is therefore linear near mid-rapidity.
Furthermore, a saturation is observed near projectile and target rapidities,
resulting in a typical $S$-shape.
\par
From the mean transverse momentum distibutions one can extract two main
observables sensitive to the EOS. One of them is the mean transverse
momentum averaged for positive values of rapidity
$<P_{x}>_{y>0}$.
A somehow equivalent observable is the transverse flow $F=\frac{d<P_{x}>}
{dY}$,
i.e. the slope
of the momentum distribution at midrapidity. It is a measure of the amount
of collective transverse
momentum transfer in the reaction i.e. intensity of the nuclear interactions.
Technically $F$ is obtained by fitting
the central part of the dependence of
$<P_{x}\hspace{0.01cm}^{\prime}(Y)>$ on $Y$
with a sum of first and third order polynomial function.
 The coefficient of the first order term is the flow $F$.
The fit was done for Y between 0.4 $\div$ 1.9 for C-Ne and 0.2 $\div$ 2 for C-Cu.
The straight line in Fig. 5 shows the result of this fit for
the experimental data in C-Ne.
The values of $F$ corrected by $k_{corr}$ are: $F$=134$\pm$12 (MeV/c)
for C-Ne and
$F$=198$\pm$13  (MeV/c) for C-Cu. The influence of the admixture
of ambiguously identified $\pi^{+}$ mesons  on the results
have been analysed.
 The error in flow $F$ includes the statistical and systematical
 errors.
The $F$ increases with atomic number of target $A_{T}$, which indicates on the
rise of collective directed flow effect.
The Quark Gluon String Model 
(QGSM) was used for a comparison with our experimental data.
In the generator COLLI which is based on the QGSM, there are two 
possibilities to generate events:
1) at not fixed impact parameter $\tilde{b}$ and 2) at fixed $b$.
The experimental
and QGSM results coincide within the errors.  For generated events
the component in the true reaction plane $ P_{x}$ had been calculated.
In the model calculations the reaction plane is known a priori and is referred as
the true reaction plane.
The dependences of  $<P_{x}(Y)>$ on $Y$ for 
not fixed impact parameter $\tilde{b}$ and at fixed $b$=2.20 fm.
 are shown on
Fig. 5. For the visual presentation,
we approximated these dependences by polynoms (the curves on Fig. 5).
From the comparison of the dependences  of $<P_{x}(Y)>$ on $Y$ obtained
by the model in  two regimes - for fixed and  not fixed $\tilde{b}$,
we conclude,
that the results are consistent and it seems, that in our experiment the values of
b=2.20 fm for C-Ne, b=2.75 fm for C-Cu are probable.
The QGSM yields a significant flow signature, which follows trends similar
to the experimental data.
 To be convinced, that the
significant sidewards deflection in Fig. 5
(for both experiment and QGSM) is
due to correlations within the events, and can not be the result of detector
biases or finite-multiplicity effects,
the $<P_{x}(Y)>$ on $Y$ has been obtained
for events composed by randomly selecting
tracks from different QGSM events (within the same multiplicity range)
(Fig. 5). One can see from Fig. 5, that in these events
there is no correlation with reaction plane.
The values of $F$, obtained
from the QGSM are $F$=95$\pm$9 (MeV/c) for C-Ne and $F$=153$\pm$13 (MeV/c)
for C-Cu.
The QGSM slightly
underestimates the flow at our energies and predicts the increase of
$F$  with atomic number of target $A_{T}$.
\par
 The comparison of our flow results for protons with flow data for
various projectile/target configurations at GSI-SIS, AGS and SPS-CERN energies
was made using the scaled flow $F_{s}=F/(A_{P}^{1/3}+A_{T}^{1/3})$.
$F_{s}$ demonstrates a common scaling behaviour for flow values from
different systems over all available energy region of 0.2 $\div$ 200.0 GeV/
nucleon.
\par
 We have obtained also the mean transverse momentum per nucleon in the
reaction plane in the forward hemisphere of the c.m. system
$<P_{x}>_{y>0}$. The  corrected (multiplied on $k_{corr}$ factor)
values of $<P_{x}>_{y>0}$ are: $<P_{x}>_{y>0}$=97$\pm$11 (MeV/c) for C-Ne
and $<P_{x}>_{y>0}$=145$\pm$18 (MeV/c) for C-Cu.
  The dependence of
$<P_{x}>_{y>0}$ on beam energy and target/projectile mass
was studied.  Our results have been compared to the results
  on central Ar-KCl, Ar-BaI$_{2}$,
Ca-Ca, Nb-Nb, Ar-Pb collisions from the Plastic Ball,
Diogene and BEVALAC streamer chamber groups.
 The $<P_{x}>_{y>0}$ rises monotonically with E$_{beam}$,
irrespective of the projectile/target configurations.
\par
In view of the strong coupling between the nucleon and pion, it is intersting
to know if pions also have a collective flow behaviour and how the pion
flow is related to the nucleon flow.
We have studied the flow effects of $\pi^{-}$ mesons. For this purpose
the reaction plane was defined for the participant protons and the transverse
momentum of each $\pi^{-}$ meson was projected onto this reaction plane.
 The data exhibit the typical S-shape behaviour as for the protons.
The values of flow $F$ for $\pi^{-}$ mesons are: for C-Ne collisions
$F=29 \pm 5$ MeV; C-Cu -- $F=-47 \pm 6$ MeV. 
In C-Ne interactions the directed (sideward) flow
of $\pi^{-}$ mesons is in the same direction as that of the protons, while
in C-Cu collisions pions show the antiflow behaviour.
The anticorrelation of nucleons and pions was explained by  S.Bass et al. as due to
multiple $\pi$N scattering. However B.A. Li et al.  show, that
anticorrelation is a manifestation of the nuclear shadowing effect of
the target and projectile spectators through both pion rescattering and
reabsorptions. In our opinion, our results indicate that the flow
behaviour of $\pi^{-}$ mesons in a light sysytem C-Ne is due to the flow
of $\Delta$ resonances, whereas the antiflow behaviour in a heavier C-Cu
system is the result of the nuclear shadowing effect.
\par
The preferential emission of partciles in the direction perpendicular
to the reaction plane (i.e. "squeeze-out") is particularly interesting
since it is the only way the nuclear matter might escape without
being rescattered by spectator remnants  of the projectile and the target,
and is expected to provide direct information on the hot and dense
participant region formed in high energy nucleus-nucleus interactions.
This phenomenon was predicted by hydrodynamical calculations, and first
has been confirmed for charged particles by the Plastic Ball collaboration.
 Then this effect was clearly identified in the experiments at GSI
set-ups KAOS, TAPS, FOPI,
 KAON, LAND and DIOGENE
group at SATURNE by the observation of an enhanced out-of-plane emission
of protons, neutrons, mesons and charged fragments.
\par
In order to extend these investigations, the azimuthal
angular distributions ($\phi$) of the protons and pions in C-Ne and C-Cu
collisions have been studied.
The angle $\phi$
is the angle of the transverse momentum of each particle in an  event with
respect to the reaction plane ($cos\phi=P_{x}/P_{t}$). 
 The analysis was
restricted only to the mid-rapidity region by applying a cut around the
center-of-mass rapidity. Fig. 6 shows distributions for
protons in C-Ne and C-Cu collisions. For visual
presentation the data of C-Cu were shifted upwards. For $\pi^{-}$ mesons
the analysis was performed from 0 to 180$^{0}$ due to lower statistics
than for protons. The azimuthal angular distributions for the protons and
pions show maxima at $\phi=90^{0}$ and 270$^{0}$ with respect to the event
plane. These maxima are associated with preferential particle  emission
perpendicular to the reaction plane (squeeze-out, or elliptic flow). Thus
a clear signature of an out-of-plane signal is evidenced. 
The QGSM yields  also a signature of the  
elliptic (squeeze-out) flow effects in C-Ne and C-Cu collisions for protons.
 Azimuthal distributions have been parametrized by a second order polynomial
function:
\begin{eqnarray}
\frac{dN}{d\phi}=a_{0}(1+a_{1}cos\phi+a_{2}cos2\phi)
\end{eqnarray}
 the parameter $a_{2}$ of the anisotropy term $a_{2}cos2\phi$ have
been extracted.
The anisotropy factor $a_{2}$ is negative for out-of-plane enhancement
(squeeze-out) and is the measure  of the strength of the anisotropic
emission.
 The ratio $R$ of the number of particles emitted in the
perpendicular direction to the number of particles emitted in the reaction
plane, which represents the magnitude of the out-of-plane emission signal
$R=(1-a_{2})/(1+a_{2})$ was also calculated.
A ratio $R$ larger than unity implies a preferred out-of-plane emission.
The $a_{2}$ and
$R$ are increasing for both protons and $\pi^{-}$ mesons, with increasing
the transverse momentum and the mass number of target $A_{T}$ and also with
narrowing of the cut applied around the center of mass rapidity. 
The elliptic flow  is more pronounced for protons than for
$\pi^{-}$ mesons.
Our results
on rapidity, mass and transverse momentum dependence of the azimuthal anisotropy
are consistent with analysis from Plastic Ball, FOPI in
Ni-Ni, Xe-CsI, Au-Au collisions from 0.15 to 1.0 GeV/nucleon for protons,
light fragments and $\pi^{\pm}$ mesons; KAOS in Au-Au at 1.0 GeV/nucleon
for protons, light fragments, pions and kaons; KAON in Bi-Bi collisions
at 0.4, 0.7 and 1.0 GeV/nucleon for $\pi^{\pm}$ mesons; TAPS collaboration
in Au-Au collisions at 1.0 GeV/nucleon for $\pi^{0}$ mesons and are confirmed
by  Isospin Quantum Molecular Dynamics model calculations. 
\par
The squeeze-out of nucleons perpendicular to the reaction plane is due
to the high compression of nuclear matter in the central hot and dense
reaction zone (it is genuinely collective effect, increasing linearly with
$A$). Since the $\Delta$ resonances are expected to flow with the nucleons
a similar anisotropy effects could be exhibited by their decay products, the pions.
According to the microscopic transport model BUU, the mechanism of the
azimuthal anisotropy of pions is found to be the shadowing effect of the
spectator matter through  both pion reabsorptions and rescatterings.
\par
In experiments at AGS and at CERN SPS the elliptic
flow is typically studied at midrapidity and is quantified in terms of
the second Fourier coefficient $v_{2}=<cos2\phi>$, which is related to
$a_{2}$ via $v_{2}=a_{2}/2$ and measures the elliptic flow. 
We have estimated $v_{2}$ 
both for C-Ne and C-Cu. The obtained results of the  elliptic flow
excitation function for protons in C-Ne and C-Cu collisions 
has been compared with the data
 in the available energy region of 0.2 $\div$ 200.0 GeV/nucleon.
The dependence of the
elliptic flow excitation function (for protons) on energy $E_{lab}$ is
displayed in Fig. 7. 
The excitation function $v_{2}$  clearly
shows an evolution from negative ($v_{2} < 0$) to positive ($v_{2} > 0$)
elliptic flow within the region of $2.0 \leq E_{beam} \leq 8.0$ GeV/nucleon
i.e. the transition from out of plane
enhancement to preferential in-plane emission (Fig. 7)
and point to an apparent transition energy $E_{tr} \sim 4$ GeV/nucleon.
The elliptic flow at AGS for Au-Au
(E895 at 6.0, 8.0 GeV/nucleon and E877 at 10.8 GeV/nucleon) and at the SPS
for Pb-Pb collisions shows in-plane enhancement, both for protons and
pions in the full rapidity range.
Studies based on transport models
have indicated that the value for $E_{tr}$ depends on the nuclear EOS
at high densities. Using a relativistic Boltzman-Equation transport
model of P.Danielewicz et. al., it has been found a softening of 
the EOS from a stiff
form ($K\sim380$ MeV) for beam energies below $E_{tr}$ to a softer form
($K\sim210$ MeV) for beam energies above $E_{tr}$.
\par
 The method of differential flow analysis was suggested as 
the  complementary approach to the standard
transverse momentum analysis of P.Danielewicz and G.Odyniec and was
first used by the E877 collaboration.
Let $N^{+}(N^{-})$ be the number of particles
emitted in the same (opposite) direction of the transverse flow near projectile
rapidity, then the ratio
$R(p_{t})=(dN^{+}/dp_{t})/(dN^{-}/dp_{t})$ as a
function of $p_{t}$  is a direct measure of the strength of differential flow
near the projectile rapidity. It has been shown experimentally that more detailed
information about the collective flow can be obtained by studying this ratio. 
It was shown theoretically by Li et al.
that the ratio $R(p_t)$ at high $p_{t}$
is particularly useful in studying the EOS of dense and hot matter formed in relativistic
heavy-ion collisions.
We examine the transverse momentum spectra of protons detected on
the same $(dN^{+}/dp_{t})$ and opposite $(dN^{-}/dp_{t})$ side of the
transverse flow (reaction plane) in central C-Ne and C-Cu 
interactions, respectively. These protons are not used in the definition
of the reaction plane. Only protons emitted in the rapidity intervals
of 1.7$<$ Y $<$ 2.4 for C-Ne
and 1.3$<$ Y $<$ 2.7 for C-Cu, respectively, were considered.
The chosen rapidity ranges are around the projectile rapidity of
$Y_{proj}=2.28$. In these rapidity ranges particles with high transverse
momenta must have suffered very violent collisions and thus originate
most likely from the very hot and dense participant region.
On the other hand, particles with low transverse
momenta are mostly from cold spectators. 
 A clear excess
of protons emitted to the same side of the directed flow for both C-Ne and C-Cu collisions
has been obtained.
In both collisions the spectra show typical exponential behaviour for
p$_{t} \geq $ 0.2 GeV/c. The spectra for particles in
the same and opposite directions of the transverse flow are approximately
parallel to each other at p$_{t}$ larger than about 0.7 GeV/c. 
To study the strength of differential transverse flow, we compare
 the ratios $R(p_{t})$ as a function of p$_{t}$ for C-Ne and
C-Cu collisions.  The ratios increase gradually at low p$_{t}$
and reach a limiting value of about 1.9 and 2.5 in the reaction of C-Ne
and C-Cu, respectively. The values of $R(p_t)$ are greater than one in the
whole transverse momentum range, indicating that protons are emitted preferentially
in the flow direction at all transverse momenta. It is an
unambiguous signature of the sideward collective flow. 
These
findings
are similar to those observed by the E877 collaboration for protons in central Au-Au
collisions at a beam energy of 10.8 GeV/nucleon. The similar behaviour show
the spectra of QGSM generated data for C-Ne and C-Cu collisions and coincide with the 
experimental ones.
From the transverse momentum distributions of protons emitted in the
reaction plane to the same  and opposite  side of the transverse
flow the temperatures and flow velocity $\beta$ have been extracted
in C-Ne  and C-Cu collisions.
The observations are in qualitative
agreement with predictions of a transversely moving thermal model.
In the whole range of transverse momentum studied, pions are found to be
preferentially emitted in the same direction of the
proton transverse flow in the reaction of C-Ne, while an anti-flow of pions is found in the
reaction of C-Cu due to stronger shadowing effects of the heavier target.
The differential flow data for both protons and pions provide a more
stringent testing ground for relativistic heavy-ion reaction theories.\\
{\underline{The Main Results of the Thesis}}\\
1. Multiplicity distributions of $\pi^{-}$ mesons, characteristics
of these distributions (average multiplicity $<n_{-}>$, dispersion
$D_{n_{-}}$, etc),
average number of interacting nucleons $<Q>$ and the ratio
$R_{-}=<n_{-}>/<Q>$ have
been obtained in central C-C, F-Mg and Mg-Mg interactions at energy of
3.4 GeV/nucleon
 with use of the spectrometer GIBS.
  Values of $<Q>$ increase with the target mass number $A_{T}$.
Values of $R_{-}$ obtained in C-C, F-Mg and Mg-Mg interactions coincide
with results
obtained earlier at SKM-200 set-up for symmetric and approximately symmetric
systems
of C-Ne, O-Ne and Ne-Ne.
Values of $R_{-}$ coincide with values calculated by using
the simple thermodynamical model.
 Empirical dependence of $R_{-}$ on masses of interacting
nuclei ($A_{P}$, $A_{T}$) $R_{-}=C \cdot (A_{P}/A_{T})^{\alpha}$, obtained for
inelastic collisions at lower Bevalac (LBL) energies up to 2 GeV/nucleon,
 have been qualitatively confirmed at our energy of 3.4 GeV/nucleon.
The Dubna Cascade Model reproduces the dependence of $R_{-}$ on $A_{P}/A_{T}$.
It has been shown, that linear dependence of $R_{-}$ on the energy in c.m.
system $E_{cm}$ observed
at Bevalac for the various symmetric pairs of nuclei up to 1.8 GeV/nucleon
is continued up to our energy of 3.4 GeV/nucleon. Obtained values
of $R_{-}$ allows one to extend the region of comparison with predictions
of theory.
Obtained results on $<Q>$ and $R_{-}$ for C-C collisions agree with
results of the 2-Metre Propane Chamber collaboration for C-C at the same
energy of E=3.4 GeV/nucleon.
\par
2. A study of pion production in central He-Li, He-C, C-Ne, C-Cu, C-Pb, O-Pb
and Mg-Mg collisions was carried out.
 Average kinematical characteristics of $\pi^{-}$ mesons such as multiplicity, momentum,
transverse momentum, emission angle, rapidity and corresponding distributions have been
obtained for Mg-Mg. The Quark Gluon String Model (QGSM) satisfactorily
describes experimental results.
It has been shown that in Mg-Mg collisions the average rapidity $<Y>$ does not depend
on the multiplicity of pions, meanwhile the average transverse momentum
$<P_{T}>$ slightly decreases with multiplicity. These average kinematical characteristics
are similar to characteristics of N-N collisions at the same energy.
Results for Mg-Mg interactions coincide (agree) with our previous results for symmetric
and approximately symmetric pairs of nuclei C-C, C-Ne, O-Ne, Ne-Ne obtained
at SKM-200 set-up.
 The QGSM reproduces the dependence of $<Y>$ and $<P_{T}>$ on $n_{-}$ for Mg-Mg.
 Rapidity distributions of pions in various regions of P$_{T}$ have been investigated.
Shape of $Y$ distributions changes with increase of the transverse momentum
of $\pi^{-}$ mesons: the fraction of pions increases in the central
region and decreases in fragmentational regions of colliding nuclei.
$Y$ distributions become narrower.
Analysis of $Y$ distributions shows that central regions
of these distributions are enriched with pions of large transverse
momentum (as compared to fragmentational regions of colliding nuclei).
The QGSM reproduces $Y$ distributions in various regions of P$_{T}$ well enough.
Analysis of angular distributions of $\pi^{-}$ mesons in He-Li,
He-C, C-Ne, C-Cu, C-Pb, O-Pb and Mg-Mg collisions have been performed.
The anisotropy coefficient {\it {a}} has been obtained. It has
been shown that the parameter {\it{a}} is same for the symmetric  (or
approximately symmetric, $A_{P}$$\sim$$A_{T}$ ) system of nuclei (He-Li
and  Mg- Mg) and increases slowly with the mass number of projectile
 $A_{P}$  and  target $A_{T}$  for other pairs of
nuclei. The QGSM reproduces angular distributions of pions well enough and
values of the {\it {a}} extracted from the spectra generated by QGSM agree
with exprimental ones within the errors.
 Dependence of {\it{a}} on kinetic energy in c.m.s $E^{*}_{kin}$
 and multiplicity of pions  n$_{-}$ has been studied. Anisotropy
coefficient increases linearly with  E$^{*}$$ /$E$^{*}_{max}$ for all pairs
of nuclei. In the range of 100 MeV pions are emitted isotropically.
These results qualitatively agree with predictions of the Dubna
Intranuclear Cascade Model.
 At small multiplicities of pions ( $n_{-} \leq~ <n_{-}>$ ), the degree of
anisotropy is greater, than at high multiplicities ( $n_{-} >~ <n_{-}>$ ).
Decrease of the parameter {\it{a}} for more central events ( $n_{-} >
~<n_{-}>$ ) indicates, that angular distributions of pions become
more isotropic at small impact parameters (b$\rightarrow$0).
\par
3. Identification of protons have been carried out in central C-Ne and C-Cu
collisions.
The statistical method of identification utilized in the neural nets
method and based on the similarity of $\pi^{-}$ and $\pi^{+}$ mesons spectra
have been used in order to separate $\pi^{+}$ mesons  from protons.
After identification the admixture of $\pi^{+}$ mesons have been reduced to
(5-7)$\%$.
\par
4. Characteristics of protons in central He-Li, He-C, C-C, C-Ne,
C-Cu and C-Pb collisions have been studied.
Average values of momentum $<P>$, transverse momentum $<P_{T}>$, rapidity
$<Y>$ and emission angle $<\Theta>$ have been obtained. Experimental results
have been compared with predictions of the Quark-Gluon String Model. The
model satisfactorily describes main characteristics and the spectra of
protons, though  slightly underestimates $<P>$ and overestimates $<\Theta>$.
\par
5. Temperatures of protons  and $\pi^{-}$ mesons in He-Li, He-C, C-C, C-Ne,
 C-Cu, C-Pb, O-Pb and Mg-Mg central collisions have been determined from kinetic energy
$E_{k}$ and transverse momentum spectra. Proton temperatures
$T_{p}$ increase with mass numbers of projectile ($A_{P}$) and target
($A_{T}$) from $T_{p}=(118\pm2)$ MeV for He-Li to
$T_{p}=(141\pm2)$ MeV for C-Pb.
The proton temperature $T_{p}$  in C-Ne, C-Cu and C-Pb collisions agrees
with prediction of the Hagedorn thermodynamic model.
Experimental results have been compared with the QGSM. The QGSM agrees with
experimental temperatures of protons and reproduces dependence of $T_{p}$ on
$A_{P}$, $A_{T}$.
 For light nuclei He-Li, He-C and C-Ne, the spectra of $\pi^{-}$
mesons are described by a single exponential law and temperatures of pions
$T_{\pi^{-}}$ does not depend on $A_{P}$, $A_{T}$. Temperatures obtained
from energy spectra, are $T_{\pi^{-}} \simeq 87$ MeV and from $P_{T}$
spectra,
$T_{\pi^{-}} \simeq 95$ MeV. The QGSM describes experimental results well.
For C-Cu, C-Pb, O-Pb and Mg-Mg interactions the shape of
energy and transverse momentum spectra of $\pi^{-}$ mesons
are concave and the sum of two exponentials ($T_{1}$, $T_{2}$) is necessary
to reproduce the data. The relative
yield of the high temperature component ($T_{2}$) is $ \simeq 24\%$ for C-Cu,
C-Pb and Mg-Mg interactions. The QGSM data of C-Cu, C-Pb, O-Pb and Mg-Mg
collisions also exhibit concaved shape and two temperatures have been
obtained from spectra. Values of temperatures
extracted from the QGSM data coincide with experimental ones.
 Dependence of the $T_{\pi^{-}}$ on emission angle in c.m. system
$\Theta_{cm}$ and on rapidity $Y$ has been studied for He-Li, He-C, C-Ne,
Mg-Mg and C-Cu collisions. $T_{\pi^{-}}$ falls off from $T_{\pi^{-}}$=110
MeV at $\Theta_{cm}$=30$^{0}$   to $T_{\pi^{-}}$=90 MeV at  $\Theta_{cm}$=90$^{0}$
and then increases. Such behaviour has been interpreted as the effect of direct
"corona" production of $\Delta$'s. Rapidity dependence of the
$T_{\pi^{-}}$ shows a bell-shaped behaviour with the maximum at Y=Y$_{beam}$/2.
 Dependence on the rapidity is more
evident for $T_{2}$ than for $T_{1}$ in Mg-Mg and C-Cu collisions.
The QGSM reproduces experimental results.
Correlation of pion kinematical characteristics with $\Lambda$'s momentum
in the nucleon-nucleon (N-N) c.m. system was investigated in
Mg-Mg collisions. Events with a $\Lambda$ produced within ("$\Lambda^{in}$" events)
and beyond ("$\Lambda^{out}$" events) the
N-N c.m. kinematical limits for nucleon-nucleon collisions were considered.
Kinematical characteristics and temperatures of pions from
$\Lambda^{in}$  and $\Lambda^{out}$  events do not reveal any significant
difference when compared between and with corresponding characteristics and
temperatures of pions produced in ordinary central Mg-Mg collisions.
The QGSM reproduces experimental results and reveals trend similar to
experimental data. For C-Ne and C-Cu interactions temperatures of identified
$\pi^{+}$ mesons have been determined. One temperature have been observed
in C-Ne and two temperatures in C-Cu collisions, similarly as for
$\pi^{-}$ mesons.
\par
6. The light front variables $\zeta^{\pm}$ and $\xi^{\pm}$, which define
 the so called horospherical coordinate system in the Lobachevsky space have
been used to study inclusive spectra of $\pi^{-}$ mesons in He(Li,C), C-Ne,
C-Cu, Mg-Mg and O-Pb collisions.
 In $\zeta^{\pm}$ ($\xi^{\pm}$) distributions the points
$\tilde{\zeta^{\pm}}$ ($\tilde{\xi^{\pm}}$) have been singled out, which
divide the phase space of secondary $\pi^{-}$ mesons into two regions
with significantly different characteristics, in one of which the thermal
equilibrium seems to be reached. Separation points are the points of maxima
in corresponding $\xi^{\pm} (\zeta^{\pm}$)-spectra (or corresponding paraboloids
in the phase space).
 Characteristics of $\pi^{-}$ mesons (momentum, angular,
$p_{T}^{2}$ -- distributions) in these two regions differ significantly.
In particular, secondary pions with $|\zeta^{\pm}|>|\tilde{\zeta}^{\pm}|$ 
($|\xi^{\pm}|<|\tilde{\xi}^{\pm}|$)
have almost flat angular distribution in the centre of mass frame, whereas
pions with $|\zeta^{\pm}|<|\tilde{\zeta}^{\pm}|$
 ($|\xi^{\pm}|>|\tilde{\xi}^{\pm}|$) are produced sharply anisotropically.
Thus one can say that the problem of separation of "pionization" and fragmentation
components seems to be solved.
Corresponding temperatures  of pions $T$ are extracted through fitting of data
by the Boltzmann distribution. Dependence of the temperature on mass numbers
of projectile ($A_{P}$) and target ($A_{T}$)nuclei have been studied. The
temperature decreases with increase of $(A_{P}*A_{T})^{1/2}$.
 Experimental results have been compared with predictions of the QGSM.
The QGSM satisfactorily reproduces experimental data for light and
intermediate-mass nuclei.
 The use of light front variables can help to distinguish different
dynamical contributions, or test basic principles in other types of analysis,
 such as two-particle correlations, HBT-interferometry and transverse flow
studies.
\par
7. The transverse momentum technique have been used to analyse charged particle
exclusive data event by event in the central C-Ne and C-Cu interactions at a
momentum of 4.5 GeV/c/nucleon (E=3.7 GeV/nucleon). Results for protons and
$\pi^{-}$ mesons are presented in terms of the mean transverse momentum
projected onto the estimated $<P_{x}\hspace{0.01cm}^{\prime}>$ and corrected
$<P_{x}>$ reaction planes as a function of rapidity $Y$ and normalized
rapidity $Y/Y_{proj}$ ($Y_{proj}$ - is the rapidity of the projectile in the
laboratory system). Observed dependences of the $<P_{x}>$ on rapidity for
protons and $\pi^{-}$ mesons show typical S-shape behaviour reflecting the
presence of the directed flow effects. The directed flow effect is larger
for protons than for $\pi^{-}$ mesons. In C-Ne interactions the directed
(sideward) flow of $\pi^{-}$ mesons is in the same direction as that of
protons, while in C-Cu collisions pions show the antiflow behaviour.
From the dependence of $<P_{x}\hspace{0.01cm}^{\prime}>$ on $Y$ (respectively
$<P_{x}>$ on $Y/Y_{proj}$) we have extracted the flow $F$, defined as the
slope at midrapidity for both protons and $\pi^{-}$ mesons. It is a measure
of the amount of collective transverse momentum transfer in the reaction.
Due to uncertainties in determination of the reaction plane correction
factors $k_{corr}$ have been estimated by two methods, which yield
consistent results.
The $F$ increases with atomic number of target $A_{T}$, which indicates
on rise of collective directed flow effect.
 The Quark Gluon String Model (QGSM) was used for comparison with
 experimental data. The QGSM yields a signature of sideward (directed)
 flow effects in C-Ne and C-Cu collisions for protons.
The QGSM underestimates the value of flow $F$.
The estimated $<P_{x}\hspace{0.01cm}^{\prime}>_{y>0}$ and corrected
(multiplied on $k_{corr}$ factor) $<P_{x}>_{y>0}$ values of the mean
transverse momentum in the reaction plane in the forward hemisphere of
the c.m. system have been obtained. Values of $<P_{x}>_{y>0}$ had been
compared with results at lower energies of 0.4 - 1.8 GeV/nucleon for
various projectile/target configurations. The $<P_{x}>_{y>0}$ increases
with the beam energy. Comparison of our flow results for protons with
flow data for
various projectile/target configurations at GSI-SIS, AGS and SPS-CERN
energies was made using the scaled flow $F_{s}=F/(A_{P}^{1/3}+A_{T}^{1/3})$.
$F_{s}$ demonstrates a common scaling behaviour for flow values from
different systems over all available energy region of 0.2 $\div$ 200.0 GeV/
nucleon.
\par
8. A clear signature of elliptic flow (squeeze-out) have been obtained
from azimuthal distributions of protons and $\pi^{-}$ mesons with respect
to the reaction plane at mid-rapidity region in both, C-Ne and C-Cu collisions.
The QGSM yields also a signature of the
elliptic flow effects in C-Ne and C-Cu collisions for protons.
 Azimuthal distributions have been parametrized by a second order polynomial
function, the parameter $a_{2}$ of the anisotropy term $a_{2}cos2\phi$ have
been extracted. The ratio $R$ of the number of particles emitted in the
perpendicular direction to the number of particles emitted in the reaction
plane, which represents the magnitude of the out-of-plane emission signal
$R=(1-a_{2})/(1+a_{2})$ was also calculated. The elliptic flow  was
shown to increase with the transverse momentum, mass number of target $A_{T}$
and also with narrowing of the
rapidity range. It is more pronounced for protons than for $\pi^{-}$ mesons.
 The second Fourier coefficient $v_{2}=<cos2\phi>$, which is related to
$a_{2}$ via $v_{2}=a_{2}/2$ and measures the elliptic flow, have been
estimated both for C-Ne and C-Cu. Obtained results on the elliptic flow
excitation function for protons in C-Ne and C-Cu collisions
has been compared with data in the available energy region of 0.2 $\div$
200.0 GeV/nucleon.
The excitation function $v_{2}$ clearly shows an evolution from negative
($v_{2} < 0$) to positive ($v_{2} > 0$) elliptic flow within the region
of $2.0 \leq E_{beam} \leq 8.0$ GeV/nucleon and point to an apparent
transition energy $E_{tr} \sim 4$ GeV/nucleon.
\par
9. The differential transverse flow of protons and pions in
central C-Ne and C-Cu collisions  were measured.
 The strength of proton differential transverse flow is found to first
increase gradually and then saturate with increase of transverse momentum.
Observations are in qualitative agreement  with predictions of a transversely
moving thermal model and transport BUU model.
In the whole range of transverse momentum studied, pions are found to be
preferentially emitted in the same direction of the proton transverse flow
in the reaction of C-Ne, while an anti-flow of pions is found in the
reaction of C-Cu due to stronger shadowing effects of the heavier target.
The differential flow data for both protons and pions provide a more
stringent testing ground for relativistic heavy-ion reaction theories.\\
{\underline{Approbation of the Work.}}
 Basic results of the dissertation were reported at: Seminars in the High Energy
 Physics Institite of the  Tbilisi State University and in the  Laboratory of High Energies of
JINR/Dubna; International Conference of Advances in Nuclear Physics and Related Areas,
July 8-12, Thessaloniki, Greece, 1997; International Conference on Particle and 
Nuclear Physics, November 14-19, Cairo, Egypt, 1997;
International Symposium in Elementary Particle Physics dedicated to the memory of
 G.Chikovani, September 21-23, Tbilisi, Georgia, 1998; On the 1999 European
School of Young Scientists in High Energy Physics, 22 August - 4 September,
Casta-Papiernicka, Slovac Republic, 1999; On the Seventh International Conference in
Nucleus-Nucleus Collisions, July 3-7, Strasbourg, France, 2000. The results are
 published as 14 scientific papers. 
\par
The investigations were performed with support from the following grants:
in 1993 from  the International Science Foundation (Short term support program);
in 1994 from  the International Science Foundation (Long term support program, Grant MXP000);
in 1995 from the Joint Founding Program of the International Science Foundation and 
the Government of Georgia (Grant MXP200);   
in 1997-1998 from the Georgian Department of Sciences and Tecnologies.\\
{\underline{Publications}}\\
1. Anikina M.,...Chkhaidze L., Dzhobava T et al., "Experimental data on multiplicities in 
central collisions C+C, F+Mg, Mg+Mg at 3.7 GeV/nucleon". - JINR Rapid Communications 
[34]-89, Dubna, 1989, 12pp.\\
2. Chkhaidze L., Dzhobava T. et al., "The temperatures of protons and 
$\pi^{-}$ mesons in central nucleus-nucleus interactions at a momentum of 4.5 GeV/c
per incident nucleon". -  Z.Phys., 1992, v.54C,  p.179-183\\
3. S.Avramenko,... T.Dzhobava et al., "Mg+Mg central collisions accompanied 
by $\Lambda^{0}$ production at 4.3 GeV/c per nucleon momentum". - Yad.Fiz., 1992, v.55,
p.721-735; Sov.J.Nucl.Phys., 1992, v.55, p.400-707\\
4. Chkhaidze L. and Dzobava T. , "The study of angular distributions of $\pi^{-}$ mesons
in nucleus-nucleus interactions at a momentum of 4.5 GeV/c per nucleon". - 
J.Phys., 1995,  v.21G, p.1223-1230\\
5. Chkhaidze L. and  Djobava T., "Characteristics of $\pi^{-}$ mesons produced
in nucleus-nucleus interactions at energy of 3.7 GeV per nucleon". -
Turkish J.Phys., 1997,  v.21, p.836-844\\
6. Chkhaidze L., Djobava T., Gogiberidze G., Kharkhelauri L., "The observation
of collective effects in central C-Ne and C-Cu collisions at a momentum of 4.5 GeV/c
per nucleon". - Phys.Lett., 1997, v.411B, p.26-32\\
7. Chkhaidze L., Djobava T., Kharkhelauri L. and Mosidze M., 
"The comparison of characteristics of $\pi^{-}$
mesons produced in central Mg-Mg interactions with the quark-gluon string
 model predictions". -
Eur.Phys.J., 1998,  v.1A,  p.299-306\\
8. Anikina M., Chkhaidze L., Djobava T., Garsevanishvili V. and
Kharkhelauri L., "Light-front analysis of $\pi^{-}$ mesons produced in Mg-Mg
collisions at 4.3 A GeV/c". -  Nucl.Phys., 1998,  v.640A, p.117-128\\
9. Anikina M., Chkhaidze L.,  Djobava T., Garsevanishvili V. and
Kharkhelauri L., "The analysis of $\pi^{-}$ mesons produced in nucleus-nucleus 
collisions at a momentum of 4.5 GeV/c/nucleon in light front variables". - 
 Eur.Phys.J., 2000,  v.7A, p.139-145\\
10. Chkhaidze L., Djobava T., Kharkhelauri L., "Study of collective  matter
flow in central C-Ne and C-Cu collisions at 3.7 GeV/nucleon". - Phys.Lett.,
2000, v.479B, p.21-28\\
11. Chkhaidze L., Djobava T. and Kharkhelauri L., "The identification of
protons and $\pi^{+}$ mesons in C-Ne and C-Cu collisions at a momentum of 4.5 GeV/c
per nucleon at SKM-200-GIBS streamer chamber". - Bulletin of the Georgian Academy
of Sciences, 2001, v.164, p.271-274\\
12. Chkhaidze L., Djobava T., Kharkhelauri L., "Experimental study of
collective flow phenomena in high energy nucleus-nucleus collisions". - Phys. Elem. Part.
At. Nucl., 2002, v.34., p.393-435\\
13. Chkhaidze L., Djobava T., Kharkhelauri L. and Li.B.A., "Differential
transverse flow in central C-Ne and C-Cu collisions at 3.7 GeV/nucleon". -
Phys. Rev., 2002, v.65C, p.054903\\
14. Chkhaidze L.,  Djobava T., Garsevanishvili V. and Tevzadze Yu.,
"Light front variables in high energy hadron-hadron and nucleus-nucleus
interactions". -  Phys. Elem. Part. At. Nucl., 2003, v. 34, p.527-545
\newpage
\begin{figure}
\begin{center}
\epsfig{file=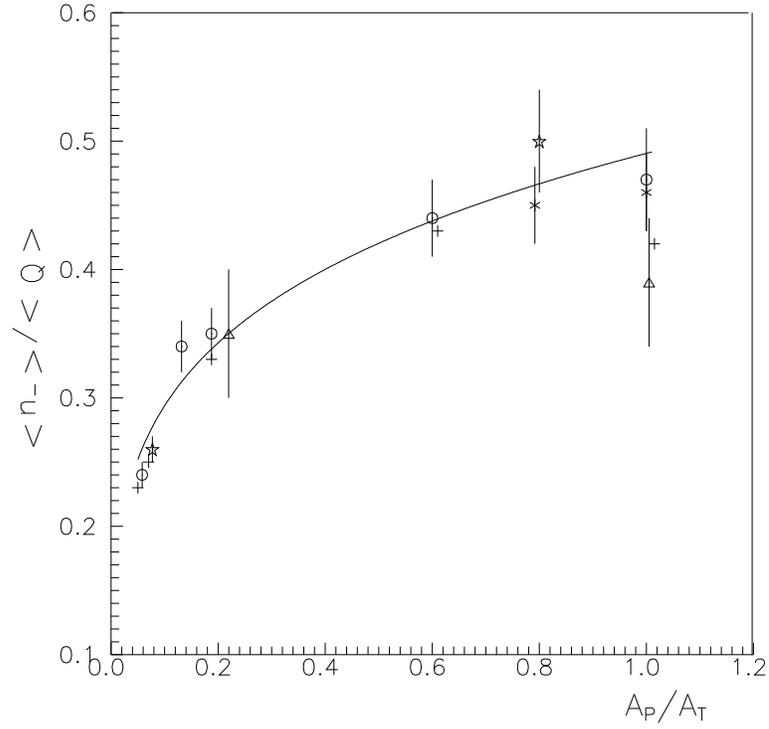,bbllx=0pt,bblly=0pt,bburx=594pt,bbury=842pt,
width=18cm,angle=0}
\end{center}
\vspace{-9.cm}
\hspace{0.cm}
\begin{minipage}{16.0cm}
\caption
{ The dependence of the
 $R_{-}=<n_{-}>/<Q>$ on the $A_{P}/A_{T}$. $\circ$ -- $C-A_{T}$,
 $\star$ -- $O-A_{T}$, $\bigtriangleup$ -- $Ne-A_{T}$,
 $\ast$ -- F-Mg, Mg-Mg.
 The data for C-C, C-Ne, C-Cu C-Zr,
C-Pb, O-Ne, O-Pb, Ne-Ne and Ne-Zr are for the trigger T(2,0).
The solid line - the result of approximation
of the data by $C \cdot (A_{P}/A_{T})^{\alpha}$. + - calculatation by
the Dubna Cascade Model (DCM).}
\end{minipage}
\end{figure}
\newpage
\begin{figure}
\begin{center}
\epsfig{file=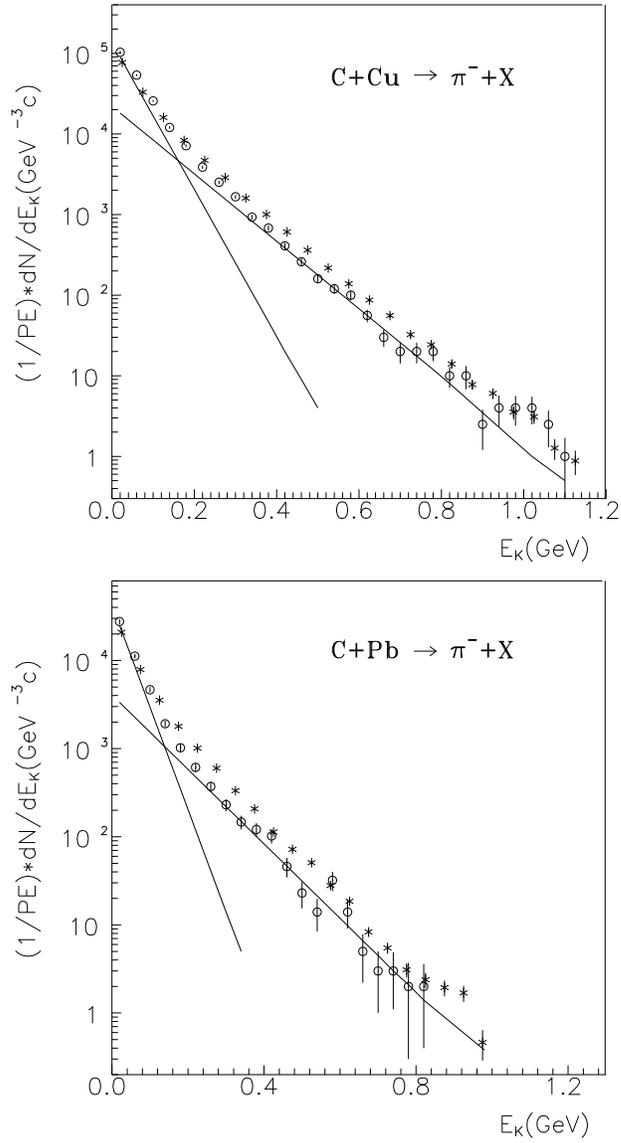,bbllx=0pt,bblly=0pt,bburx=594pt,bbury=842pt,
width=18cm,angle=0}
\end{center}
\vspace{-7.cm}
\hspace{0.cm}
\begin{minipage}{16.0cm}
\caption
{ Noninvariant kinetic energy
 spectra  of $\pi^{-}$ mesons in
 C-Cu and C-Pb collisions.
$\circ$ -- the experimental data, $\ast$ -- QGSM generated
data for fixed $b$=2.75 fm (C-Cu) and $b$=3.90 fm (C-Pb).
full lines -- the result of fitting by (5)}.
\end{minipage}
\end{figure}
\newpage
\begin{figure}
\begin{center}
\epsfig{file=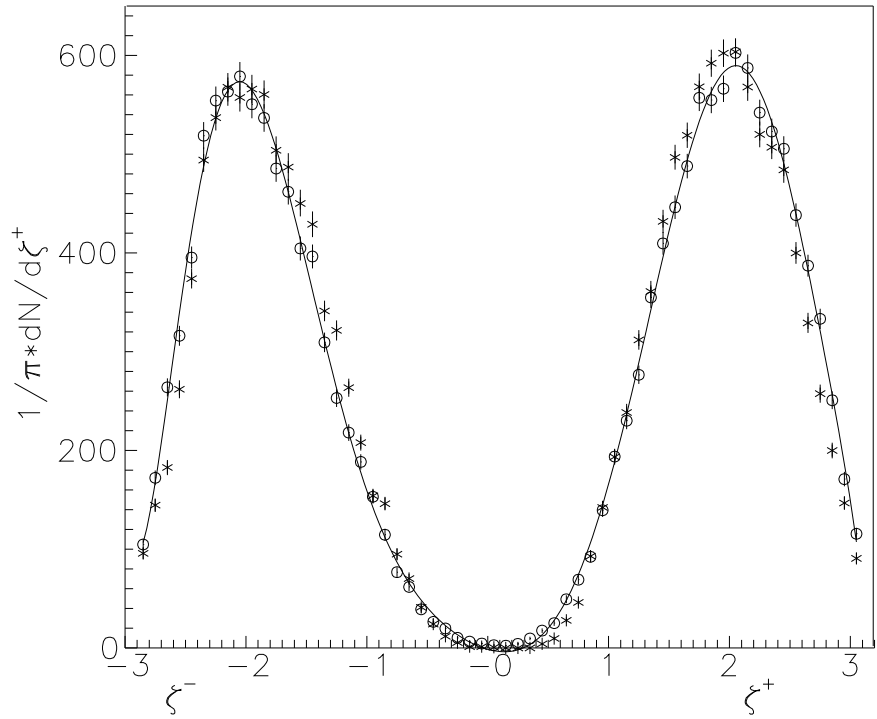,bbllx=0pt,bblly=0pt,bburx=594pt,bbury=842pt,
width=18cm,angle=0}
\end{center}
\vspace{-6.8cm}
\hspace{0.cm}
\begin{minipage}{16.0cm}
\caption
{The  $ \zeta^{\pm} $ distribution of  $\pi^{-}$ mesons from Mg-Mg interactions.
$\circ$   --  experimental data,
$\star$ -- QGSM data. The curve -- result of polinomial approximation
of the experimental data.}
\end{minipage}
\end{figure}
\newpage
\begin{figure}
\begin{center}
\epsfig{file=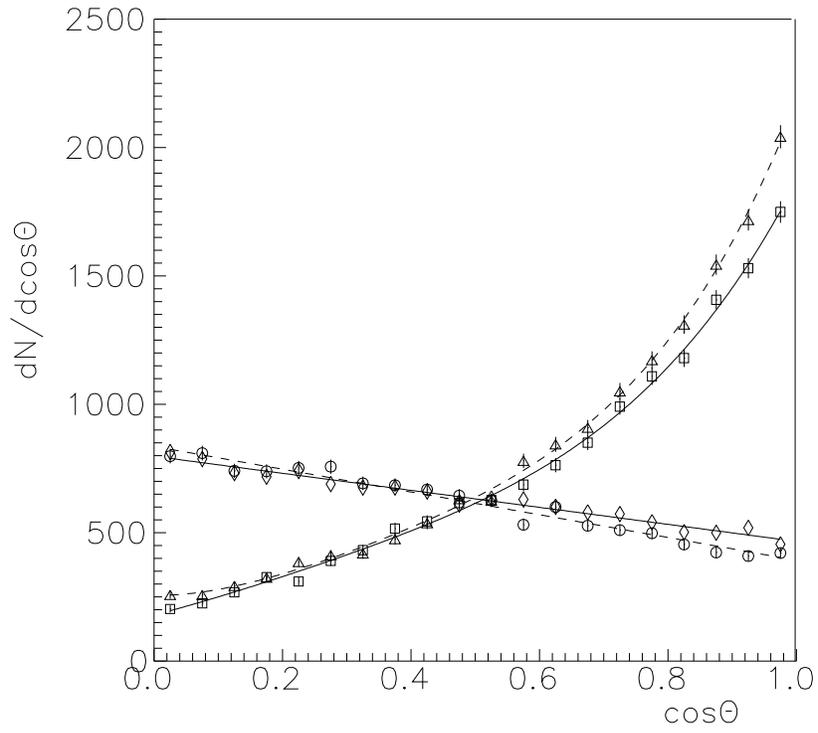,bbllx=0pt,bblly=0pt,bburx=594pt,bbury=842pt,
width=18cm,angle=0}
\end{center}
\vspace{-9.cm}
\hspace{0.cm}
\begin{minipage}{16.0cm}
\caption
{The $ cos\Theta $ distribution of  $\pi^{-}$ mesons from Mg-Mg interactions.
$\circ$   -- experimental data for $\zeta^{+} > \tilde{\zeta^{+}}$
($\tilde{\zeta^{+}}$=2.0);
$\diamond$   -- the QGSM data for $\zeta^{+} > \tilde{\zeta^{+}}$;
$\bigtriangleup$ -- experimental data for $\zeta^{+} < \tilde{\zeta^{+}}$;
squares -- the QGSM data for
$\zeta^{+} < \tilde{\zeta^{+}}$.
 Dashed lines: fit of the experimental data
by the Boltzmann distribution. Solid lines: fit of the QGSM data by
the Boltzmann distribution.}
\end{minipage}
\end{figure}
\newpage
\begin{figure}
\begin{center}
\epsfig{file=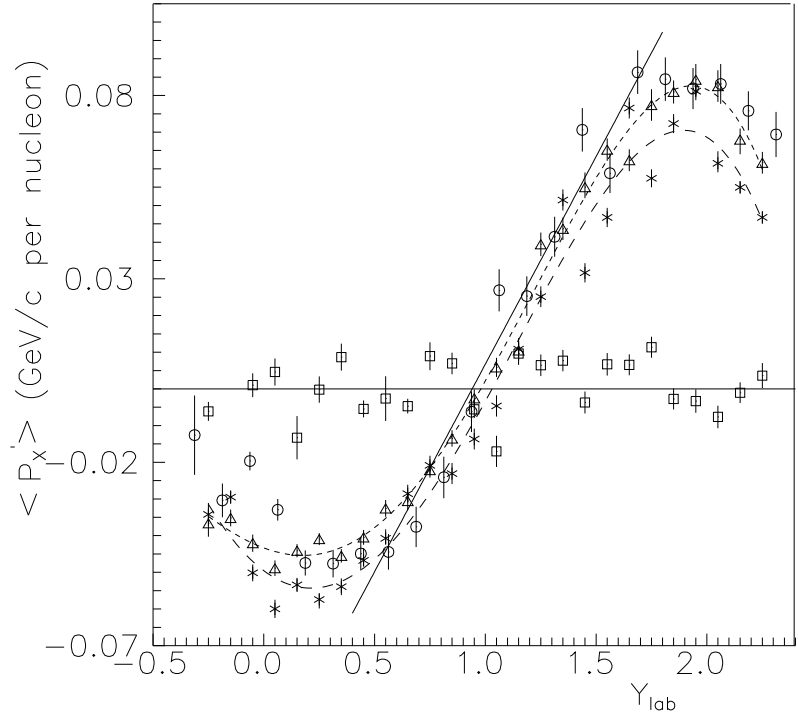,bbllx=0pt,bblly=0pt,bburx=594pt,bbury=842pt,
width=18cm,angle=0}
\end{center}
\vspace{-12.0cm}
\hspace{0.cm}
\begin{minipage}{16.cm}
\caption
{The dependence of $< P_{x}\hspace{0.01cm}^{\prime}(Y) >$ on
Y$_{Lab}$ for
protons in C-Ne collisions. $\circ$ -- the experimental data,
$\bigtriangleup$ -- QGSM generated data for fixed $b$= 2.20 fm,
$\ast$ -- QGSM generated data for not fixed $\tilde{b}$,
squares --  events composed by randomly selected
tracks from different QGSM events (within the same multiplicity range).
The solid
line is the result of the approximation of experimental data by sum of
first- and third-order polynomial function in the interval of
 Y - 0.4 $\div$ 1.9
. The dashed curves for
visual presentation of QGSM events (short dashes - for fixed $b$,
long dashes -for not fixed $\tilde{b}$)
- result of approximation by the 4-th order polynomial function.}
\end{minipage}
\end{figure}
\newpage
\begin{figure}
\begin{center}
\epsfig{file=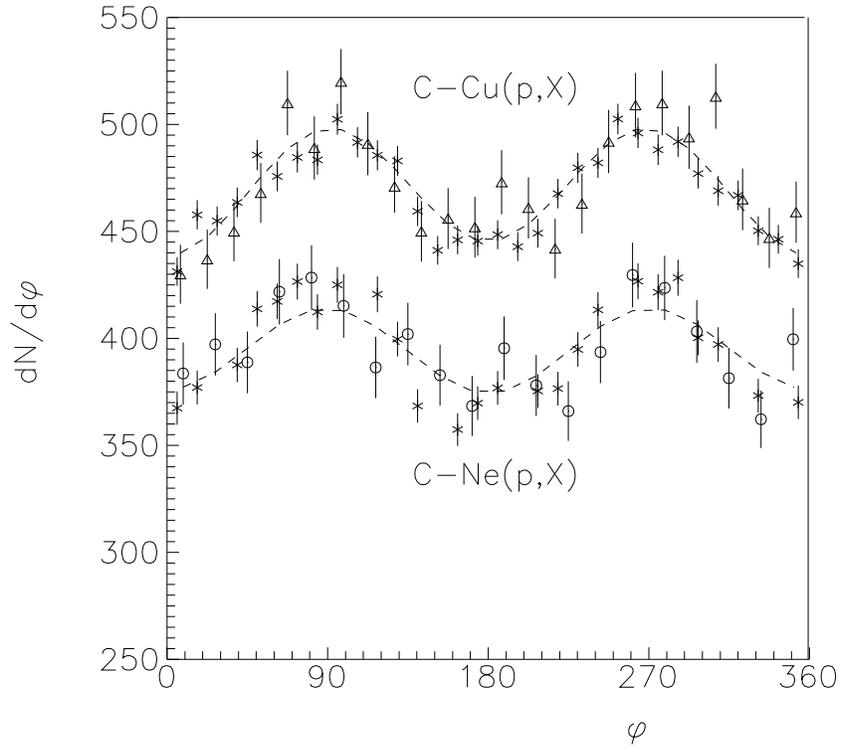,bbllx=0pt,bblly=0pt,bburx=594pt,bbury=842pt,
width=18cm,angle=0}
\end{center}
\vspace{-12.0cm}
\hspace{0.cm}
\begin{minipage}{16.0cm}
\caption
{The azimuthal distributions with respect to the reaction plane
of midrapidity
protons dN/d$\phi$ .
 $\circ$ -- for C-Ne ($-1\leq y_{cm}\leq1$),
 $\bigtriangleup$ -- for C-Cu ($-1\leq y_{cm}\leq1$) interactions,
 $\ast$ --  the QGSM data.
Also shown are the fits using the function
$dN/d\phi=a_{0}(1+a_{1}cos\phi+a_{2}cos2\phi)$.}
\end{minipage}
\end{figure}
\newpage
\begin{figure}
\begin{center}
\epsfig{file=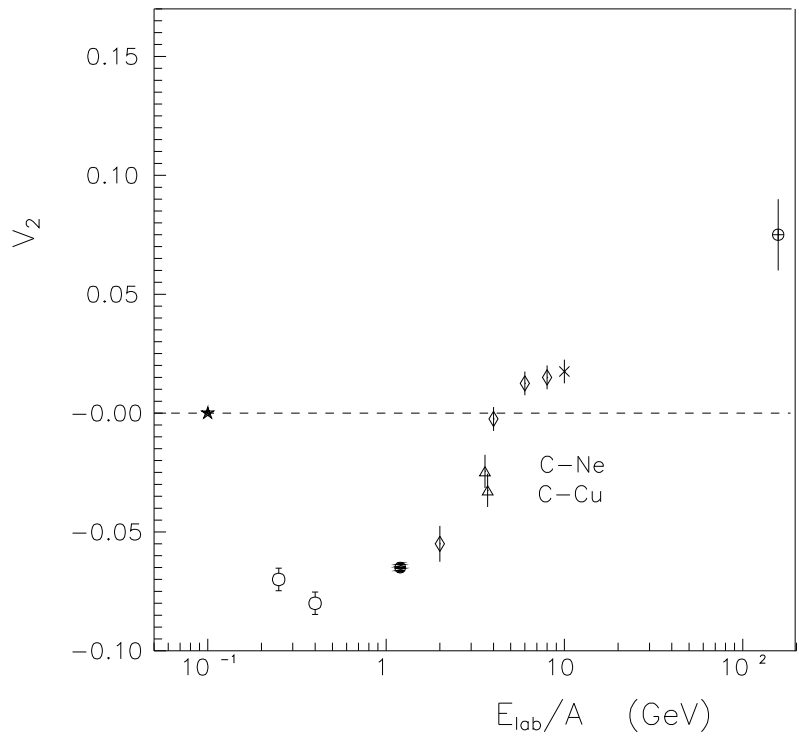,bbllx=0pt,bblly=0pt,bburx=594pt,bbury=842pt,
width=18cm,angle=0}
\end{center}
\vspace{-12.0cm}
\hspace{0.cm}
\begin{minipage}{16.0cm}
\caption
{The dependence of the Elliptic flow excitation function $v_{2}$
on energy
E$_{lab}$/A (GeV):
$\star$ -- FOPI, $\circ$ -- MINIBALL,
 $\bullet$ -- EOS,
$\diamond$
-- E-895,
$\ast$ -- E-877,
$\oplus$
-- NA49,
$\bigtriangleup$ -- C-Ne, C-Cu (our results).}
\end{minipage}
\end{figure}
\end{document}